\documentclass{aa}  

\usepackage{graphicx}
\usepackage{amsmath}
\usepackage{graphicx}
\usepackage[english]{babel}
\usepackage{subfig}
\usepackage{amssymb}
\usepackage{wrapfig}
\usepackage{pdflscape}
\usepackage{lscape}
\usepackage{longtable}
\usepackage{appendix}
\usepackage{textcomp}
\usepackage{multirow}
\bibpunct{(}{)}{;}{a}{}{,}
\usepackage{array}
\usepackage{txfonts}

\begin{document} 

   \title{Statistical multifrequency study of narrow-line Seyfert 1 galaxies \thanks{Tables 1, 2, 5, 7, 8, 12, and 13, and Figures 2, 4, 5, and 7 are only available in electronic form via http://www.edpsciences.org}}


   \author{E. J\"{a}rvel\"{a}
          \inst{1}\fnmsep\inst{2}\thanks{\email{emilia.jarvela@aalto.fi}}
          \and
          A. L\"{a}hteenm\"{a}ki\inst{1}\fnmsep\inst{2}
          \and J. Le\'{o}n-Tavares\inst{3}\fnmsep\inst{4}
          }

   \institute{Aalto University Mets\"{a}hovi Radio Observatory, Mets\"{a}hovintie 114, Kylm\"{a}l\"{a}, 02540, Finland
         \and
             Aalto University Department of Radio Science and Engineering, 13000, FI-00076 AALTO, Finland
         \and
             Instituto Nacional de Astrof\'{\i}sica \'Optica y Electr\'onica (INAOE), Apartado Postal 51 y 216, 72000 Puebla, M\'exico
         \and
             Finnish Centre for  Astronomy with ESO (FINCA), University of Turku,V\"ais\"al\"antie 20, FI-21500  Piikki\"o, Finland
             }

   \date{Received ; accepted }


  \abstract
   {High-energy $\gamma$-rays, which are produced by powerful relativistic jets, are usually associated
   with blazars and radio galaxies. In the current active galactic nuclei (AGN) paradigm, such jets 
   are almost exclusively launched from massive elliptical galaxies. Recently, however, 
   {\it Fermi}/LAT detected $\gamma$-rays from a few narrow-line Seyfert 1 galaxies and thus 
   confirmed the presence of relativistic jets in them. Since NLS1 galaxies are assumed to be 
   young evolving AGN, they offer a unique opportunity to study the production of relativistic jets in late-type galaxies.}
   {Our aim is to estimate by which processes the emission of various kinds is produced in NLS1 galaxies and to study how emission properties are connected to other intrinsic AGN properties.}
   {We have compiled the so far largest multiwavelength database of NLS1 sources. This allowed us to explore correlations 
    between different wavebands and source properties using, for example, Pearson and Spearman correlations and principal component analysis. We did this separately for radio-loud and radio-quiet sources.}
   {Multiwavelength correlations suggest that radio-loud sources host relativistic jets that are the predominant sources of 
   radio, optical, and X-ray emission. The origin of infrared emission remains unclear. Radio-quiet sources do not host 
   a jet, or the jet is very weak. In them, radio and infrared emission is more likely generated via star
   formation processes, and the optical and X-ray emission originate in the inner parts of the AGN. 
   We also find that the black hole mass correlates significantly with radio loudness, which suggests that NLS1 galaxies with more 
   massive black holes are more likely to be able to launch powerful relativistic jets.} 
   {}


   \maketitle


\section{Introduction}
\label{intro}

Narrow-line Seyfert 1 galaxies (NLS1) are a subclass of Seyfert galaxies first described in 1985 
by \citet{1985osterbrock1}. They are mostly hosted by spiral galaxies; however, a few of them are found in peculiar, interacting, or E/S0 systems \citep{2007ohta1}. 
Optical studies suggest that the disk-like host galaxies of NLS1 galaxies are more often barred (85 $\pm$ 7 $\%$) than are the disk-like hosts of broad-line Seyfert 1 galaxies (BLS1) 
(40 - 70 $\%$) \citep{2007ohta1}. 

NLS1 galaxies are believed to be rather young active galactic nuclei (AGN) in the early stages of their evolution \citep{2001mathur1}. They 
harbor low- or intermediate-mass black holes ($M_{\text{BH}} < 10^{8}$ $M_{\sun}$) \citep{2000peterson1} accreting at high rates 
(0.1-1 Eddington rate or even above) \citep{1992boroson1} and tend to lie below the normal $M_{\text{BH}}$ -- $\sigma_{\ast}$ 
(stellar velocity dispersion of the bulge) and $M_{\text{BH}}$ -- $L_{\text{bulge}}$ (luminosity of the bulge) 
relations \citep{2001mathur1}, suggesting that they are still evolving. Radio-quiet NLS1 galaxies also show enhanced star formation \citep{2010sani1}.

In NLS1 galaxies permitted emission lines are narrow, making them of comparable width with narrow forbidden lines. Characterizing spectral 
features are FWHM(H$\beta$) $< 2000$ km s$^{-1}$ \citep{1989goodrich1} and [O III]/H$\beta <$ 3 (with exceptions allowed if there are
strong [Fe VIII] and [Fe X] present) \citep{1985osterbrock1}. Some but not all NLS1 sources show strong Fe II emission \citep{1985osterbrock1}.

NLS1 sources show a strong soft X-ray excess, and some of them also exhibit very rapid, high-amplitude variability at X-rays. 
Overall they have more diverse soft X-ray (0.1-2.5 keV) photon indices ($\Gamma \approx$ 1 - 5) than Type 1 Seyfert 
galaxies ($\Gamma \approx$ 2) \citep{1996boller1}. There is a relation between FWHM(H$\beta$) and the X-ray
spectral slope $\alpha_{\text{X}}$. Sources with narrower H$\beta$ tend to have steeper X-ray spectra \citep{1992puchnarewicz1, 1996boller1}.

NLS1 sources are generally radio-quiet, but studies have shown that $\sim$7$\%$ of them are radio-loud \citep{2006komossa1}. However, they 
generally have a very compact radio morphology; evidence of large scale structures has been found only in six NLS1 
sources \citep{2010gliozzi1, 2012doi1}. All of these sources are radio-loud and have on average more massive black holes than the NLS1 population in general.
Five of six sources have $M_{\text{BH}} > 10^7 $$M_{\sun}$ \citep{2012doi1}.

The spectral energy distributions (SEDs) of some radio-loud NLS1 sources are similar to the SEDs of blazars. In them the most prominent features 
are two broad components, one extending from radio to soft X-rays and the other covering hard X-rays and $\gamma$-rays. 
The first bump is believed to be due to synchrotron emission and the second bump due to inverse-Compton (IC) scattering.

Although some NLS1 galaxies show blazar-like behavior, no $\gamma$-ray emission was expected due to their host galaxy type. That changed in 
2008 when the Large Area Telescope{\footnote{http://fermi.gsfc.nasa.gov/science/instruments/lat.html}} onboard
{\it Fermi Gamma-ray Space Telescope}{\footnote{http://fermi.gsfc.nasa.gov/}} (hereafter {\it Fermi}) detected $\gamma$-ray emission from the 
source PMN J0948+0022 identified as a NLS1 galaxy \citep{2009abdo2}.  Consequently two multiwavelength campaigns were launched 
to better understand its nature. During the first campaign (March-July 2009) it was confirmed that the $\gamma$-ray emission is indeed associated with 
the known radio-loud NLS1 PMN J0948+0022 \citep{2009abdo1, 2008yuan1}. Later, during the second multiwavelength campaign (July-September 2010) 
it flared at $\gamma$-rays and reached the extreme power of $\sim$10$^{48}$ erg s$^{-1}$ in the 0.1-100 GeV band \citep{2011foschini2}. 

So far five NLS1 sources have been detected with high significance at $\gamma$-rays confirming that they are a new class of 
$\gamma$-ray emitting AGN, in addition to blazars and radio galaxies. The discovery of $\gamma$-ray emission 
from NLS1 galaxies was remarkable because in the current AGN paradigm
powerful relativistic jets are almost exclusively launched from massive elliptical galaxies with supermassive black holes \citep{2003urry1}.
Blazars and radio galaxies, and NLS1 galaxies have different hosts (late-type in NLS1 galaxies), $M_{\text{BH}}$ (smaller in NLS1 galaxies),
accretion rates (higher in NLS1 galaxies), and radio morphologies (compact in NLS1 galaxies). Yet we now know that NLS1 galaxies can form 
and launch a fully developed relativistic jet. This poses many interesting questions concerning AGN and relativistic jet evolution, 
and challenges our current knowledge of jet systems. Therefore, NLS1 galaxies offer a great opportunity to study the evolution of 
relativistic jets and further our understanding about the mechanisms that drive AGN activity.

In this study our aim is to estimate via which processes and, if possible, where the various kinds of emission are produced in NLS1 galaxies. We address 
this issue by compiling multiwavelength observations from literature for a large sample of NLS1 sources. This allows us to explore 
correlations between different wavebands, and identify the most likely radiation mechanism responsible for the bulk of the energy 
released in NLS1 sources. We are also interested in how the emission properties are connected with other 
intrinsic AGN properties, for example, the black hole mass.

Throughout the paper we assume a cosmology with H$_{0}$ = 73 km s$^{-1}$ Mpc$^{-1}$, $\Omega_{\text{matter}}$ = 0.27 and $\Omega_{\text{vacuum}}$ = 0.73. 


\section{Sample selection}
\label{sample}

Our sample was selected using three references: \citet{2006zhou1}, \citet{2008yuan1}, and \citet{2006komossa1}. The original sample in \citet{2006zhou1}
consists of 2011 NLS1 sources selected from Sloan Digital Sky Survey{\footnote{www.sdss.org}} (SDSS) Data Release 3 with restrictions $z \lesssim$ 0.8 and the 'broad' component 
of H$\beta$ or H$\alpha$ narrower than 2200 km s$^{-1}$ at the 10 $\sigma$ or higher confidence level. From this sample, using ASI Science Data Center's 
(ASDC{\footnote{www.asdc.asi.it}}) Sky Explorer and SED Builder, we selected sources which had radio data from the Very Large Array (VLA) Faint Images of 
the Radio Sky at Twenty-Centimeters (FIRST{\footnote{www.sundog.stsci.edu}}) survey. This makes a total of 280 sources which we included in our sample. 

The sample in \citet{2008yuan1} consists of 23 radio-loud NLS1 sources selected from SDSS Data Release 5 
under the same restrictions as in \citet{2006zhou1}. 12 of them overlap with the \citet{2006zhou1} sample; we included the remaining 11 new sources to our sample. 

Our third reference \citet{2006komossa1} has a sample of 11 radio-loud NLS1 sources found by cross-correlating the Catalogue of Quasars and Active 
Nuclei \citep{2003veroncetty1} with several radio and optical catalogs using the cross-matcher application developed within the German
Astrophysical Virtual Observatory (GAVO){\footnote{www.g-vo.org}} project. The sample was limited by 
the requirement H$\beta <$ 2000 km s$^{-1}$. From \citet{2006komossa1} we were able to 
include only one source to our sample since some of them overlapped sources from \citet{2006zhou1} and \citet{2008yuan1}, and some of them did not have radio data.
Our final sample consists of 292 NLS1 galaxies.

All the data for our sample were gathered from publicly available archival sources, and it should be noted that they are therefore
not simultaneous. Studying correlations between wavebands of variable sources, such as AGN, should ideally be performed with
data that are no more than a couple of weeks apart, but in practise this is often impossible. Radio data from FIRST, optical data from SDSS, 
and X-ray data from ROSAT{\footnote{http://heasarc.gsfc.nasa.gov/docs/rosat/rosgof.html}} All Sky Survey 
(RASS) were retrieved from ASDC. We have radio and optical data for all of our sources, and X-ray data for 109 sources (hereafter called the X-ray sample). 
Data obtained from ASDC were already corrected for galactic extinction.

Infrared data from Wide-field Infrared Survey Explorer (WISE{\footnote{www.nasa.gov/wise}}) All-Sky Source Catalog{\footnote{http://wise2.ipac.caltech.edu/docs/release/allsky/}} \citep{2012cutri1}
were retrieved from the NASA/IPAC Infrared Science Archive (IRSA{\footnote{http://www.irsa.ipac.caltech.edu}}). We used a search radius of 2$\farcs$4 \citep{2011massaro1} 
and a minimum signal-to-noise ratio $>$7 in at least one band. We found a match for 291 sources. Nine sources were excluded from the sample due to the contamination 
risk{\footnote{For sources with non-zero cc$\_$flags -column it is adviced in the WISE Explanatory Supplement that the observation should 
be ignored because one or more bands might be somehow contaminated.}}, thus we have infrared data for 282 sources.
Infrared data are not corrected for the galactic absorption because the correction would be smaller than the uncertainties in magnitudes.
Flux densities (in Janskys) were computed using the WISE magnitudes{\footnote{http://wise2.ipac.caltech.edu/docs/release/allsky/expsup/sec4$\_$4h.html}}.

Additional W4-band correction suggested by the WISE Explanatory Supplement was done for 'red' sources with $\alpha>$ 1 
($\alpha$ being power-law index: $F_\nu \propto \nu ^{-\alpha} $). $\alpha$ was estimated using [W1-W2] and [W2-W3] 
colors and a table in \citet{2010wright1}. Corrections were computed with the equation: $F_{\nu}$ [W4] = 0.9 $\times$ $F_{\nu}$ [W4].
Flux densities for all of our sources are listed in Online Table~\ref{tab:bigtable2}.

Luminosities for all wavebands were computed from the equation

\begin{equation} L = 4 \pi D_{\text{L}}^2 \nu F_{\nu} \end{equation}

where $D_{\text{L}}$ is the luminosity distance to the source and $F_{\nu}$ flux density in Janskys. $D_{\text{L}}$ values were obtained from the NASA/IPAC Extragalactic
Database{\footnote{www.http://ned.ipac.caltech.edu/}} (NED).

\onllongtab{

\begin{longtab}
\begin{landscape}
\begin{longtable}{llllllllllll}

\caption{Flux densities of our whole sample.} \label{tab:bigtable2}\\
\hline\hline
\multicolumn{1}{c}{\textbf{Source}} & \textbf{FIRST} & \textbf{W1} & \textbf{W2} & \textbf{W3} & \textbf{W4} & \textbf{SDSS $u$} & \textbf{SDSS $g$} & \textbf{SDSS $r$} & \textbf{SDSS $i$} & \textbf{SDSS $z$}  & \textbf{RASS}\\

                &    (mJy)       &       (mJy) & (mJy)       &  (mJy)      &(mJy)        &  (mJy)            & (mJy)             &   (mJy)           &(mJy)              & (mJy)              & (mJy) \\

\hline
\endfirsthead
\multicolumn{10}{c}%
{\tablename\ \thetable\ -- \textit{Continued from previous page}} \\
\hline\hline
\multicolumn{1}{c}{\textbf{Source}}& \textbf{FIRST} & \textbf{W1} & \textbf{W2} & \textbf{W3} & \textbf{W4} & \textbf{SDSS $u$} & \textbf{SDSS $g$} & \textbf{SDSS $r$} & \textbf{SDSS $i$} & \textbf{SDSS $z$}  & \textbf{RASS}\\
               &      (mJy)     & (mJy)       & (mJy)       &  (mJy)      &(mJy)        &  (mJy)            & (mJy)             &   (mJy)           &    (mJy)          & (mJy)              & (mJy) \\

\hline
\endhead
\hline \multicolumn{12}{r}{\textit{Continued on next page}} \\
\endfoot
\hline
\endlastfoot

	SDSS	J002249.22-103956.2	&	1.88	&	1.73	&	2.94	&	7.22	&	16.44	&	0.26	&	0.35	&	0.37	&	0.41	&	0.53	&			\\
	SDSS	J002305.04-010743.4	&	10.56	&	0.60	&	0.60	&	2.29	&	3.15	&	0.16	&	0.24	&	0.35	&	0.46	&	0.52	&			\\
	SDSS	J002752.40+002615.7	&	4.59	&	1.92	&	2.68	&	7.61	&	17.61	&	0.12	&	0.16	&	0.22	&	0.33	&	0.33	&			\\
	SDSS	J013521.68-004402.1	&	1.99	&		&		&		&		&	0.23	&	0.52	&	0.83	&	1.23	&	1.49	&	0.0001	\\	
	SDSS	J014019.06-092110.5	&	2.12	&	1.81	&	1.91	&	10.29	&	12.81	&	0.34	&	0.76	&	1.26	&	1.81	&	2.20	&	0.00032	\\	
	SDSS	J014644.83-004043.2	&	4.33	&	4.06	&	3.94	&	19.07	&	42.52	&	0.49	&	1.06	&	1.67	&	2.47	&	2.65	&			\\
	SDSS	J022347.48-083655.6	&	1.04	&	1.53	&	2.08	&	3.83	&	11.16	&	0.11	&	0.13	&	0.18	&	0.17	&	0.25	&	0.0001	\\	
	SDSS	J022923.44-000047.9	&	2.43	&	0.74	&	1.11	&	2.26	&	7.23	&	0.09	&	0.12	&	0.12	&	0.15	&	0.15	&			\\
\end{longtable}
\end{landscape}
\end{longtab}
}

\section{Data analysis}
\label{analysis}

\subsection{Radio-loudness}
\label{sec:radioloudness}

We computed radio loudness ($RL$) for all of our sources. We used the commonly defined $RL$ value; the ratio of 1.4 GHz radio 
flux density ($F_{\text{R}}$) and 440 nm optical flux density ($F_{\text{O}}$).
For $F_{\text{R}}$ we used the K-corrected (we did K-correction as suggested in \citet{2011foschini1}) radio flux density from the FIRST 
survey and for $F_{\text{O}}$ the K-corrected B-band optical flux density calculated using SDSS $u$- and $g$-band 
magnitudes{\footnote{http://www.sdss.org/dr5/algorithms/sdssUBVRITransform.html}}$^,${\footnote{http://www.sdss.org/dr5/algorithms/fluxcal.html$\#$sdss2ab}}.
The radio loudness values for our sources are listed in Online Table~\ref{tab:bigtable1}.

\onllongtab{

\begin{longtab}
\begin{landscape}
\begin{longtable}{llllllllll}

\caption{Summary of the properties of our whole sample. In the references column Z = \citet{2006zhou1}, Y = \citet{2008yuan1} and K = \citet{2006komossa1}.} \label{tab:bigtable1}\\
\hline\hline
\multicolumn{1}{c}{\textbf{Source}}& \textbf{ref} & \textbf{RA} & \textbf{Dec} & \textbf{$z$} & \textbf{D$_\text{L}$} & \textbf{$RL$} & \textbf{$\log \lambda$L (5100\AA{})} & \textbf{FWHM(H$\beta$)}  & \textbf{$\log M_{\text{BH}}$}\\
                &              & (\textdegree) &(\textdegree)   &              &(Mpc)             &             & (ergs s$^{-1}$)                     &(km s$^{-1}$)            &($M_{\sun}$) \\

\hline
\endfirsthead
\multicolumn{10}{c}%
{\tablename\ \thetable\ -- \textit{Continued from previous page}} \\

\hline\hline

\multicolumn{1}{c}{\textbf{Source}} & \textbf{ref} & \textbf{RA} & \textbf{Dec} & \textbf{$z$} & \textbf{D$_\text{L}$} & \textbf{$RL$} & \textbf{$\log \lambda$L (5100\AA{})} & \textbf{FWHM(H$\beta$)} & \textbf{$\log M_{\text{BH}}$}\\

                &              & (\textdegree) &(\textdegree)   &              &(Mpc)             &             & (ergs s$^{-1}$)                     &(km s$^{-1}$)            &($M_{\sun}$) \\
\hline
\endhead
\hline \multicolumn{10}{r}{\textit{Continued on next page}} \\
\endfoot
\hline
\endlastfoot

SDSS	J002249.22-103956.2	&		Z	&	5.7051	&	-10.6656	&	0.4139	&	2185	&	5.61	&	44.75	&	1918	&	7.69	\\	
SDSS	J002305.04-010743.4	&		Z	&	5.771	&	-1.1287	&	0.1664	&	763	&	45.25	&	43.67	&	1214	&	6.60	\\	
SDSS	J002752.40+002615.7	&		Z	&	6.9683	&	0.4377	&	0.2053	&	966	&	29.33	&	43.88	&	1862	&	7.11	\\	
SDSS	J013521.68-004402.1	&		Z	&	23.8403	&	-0.7339	&	0.0985	&	430	&	3.77	&	43.54	&	1767	&	6.84	\\	
SDSS	J014019.06-092110.5	&		Z	&	25.0794	&	-9.3529	&	0.1353	&	609	&	2.92	&	43.72	&	1576	&	6.86	\\	
SDSS	J014644.83-004043.2	&		Z	&	26.6868	&	-0.6787	&	0.0827	&	358	&	3.92	&	43.47	&	1255	&	6.50	\\	
SDSS	J022347.48-083655.6	&		Z	&	35.9479	&	-8.6154	&	0.2608	&	1272	&	7.86	&	44.01	&	1896	&	7.21	\\	
SDSS	J022923.44-000047.9	&		Z	&	37.3477	&	-0.0133	&	0.5581	&	3132	&	22.70	&	44.67	&	1298	&	7.30	\\
\end{longtable}
\end{landscape}
\end{longtab}
}

We then divided the sources to four subsamples by their radio loudness. The subsamples are: radio-quiet (RQ; $RL$ $<$ 10, 97 sources),
radio-loud (RL; $RL$ $>$ 10, 195 sources), very radio-loud (VRL; $RL$ $>$ 100, 51 sources), and super radio-loud (SRL; $RL >$ 1000, 10 sources).
The radio-loud subsample includes very radio-loud and super radio-loud subsamples. The subsamples and their sizes by waveband are presented in Table~\ref{tab:subsamples}.

\begin{table*}[ht]
\caption{Sample sizes at different wavebands, mean redshifts and their standard deviations (std), and mean black hole masses and their standard deviations for the whole sample and the subsamples.}
\centering
{
\begin{tabular}{l l l l l l l l l l l}
\hline\hline
sample & $N$   & $N_\text{R}$ & $N_{\text{IR}}$ & $N_\text{O}$ & $N_\text{X}$ & $N_{\text{IR}}$ and $N_\text{X}$ & \multicolumn{2}{c}{$z$} & \multicolumn{2}{c}{$\log M_{\text{BH}} (M_{\sun})$}\\ 
       &     &       &          &       &       &                    & mean     & std        & mean   & std    \\ \hline
all    & 292 & 292   & 282      & 292   & 109   & 105                & 0.35    & 0.20      & 7.10 & 0.47 \\ 
RQ     & 97  & 97    & 90       & 97    & 60    & 56                 & 0.22    & 0.17      & 6.96 & 0.52 \\ 
RL     & 195 & 195   & 192      & 195   & 49    & 49                 & 0.41    & 0.19      & 7.18 & 0.43 \\ 
VRL    & 51  & 51    & 51       & 51    & 16    & 16                 & 0.48    & 0.17      & 7.31 & 0.39 \\ 
SRL    & 10  & 10    & 10       & 10    & 4     & 4                  & 0.54    & 0.17      & 7.30 & 0.40 \\ \hline
\end{tabular}
}
\label{tab:subsamples}
\end{table*}

\subsection{Parent population}
\label{sec:parentpop}

We used the two-sample Kolmogorov-Smirnov test (hereafter K-S test) with 5$\%$ significance level
(probability value p$<$0.05) in order to examine whether the parent population of our radio-quiet and radio-loud subsamples is the same. 

\begin{figure}[ht!]
\centering
  \includegraphics[width=0.49\textwidth]{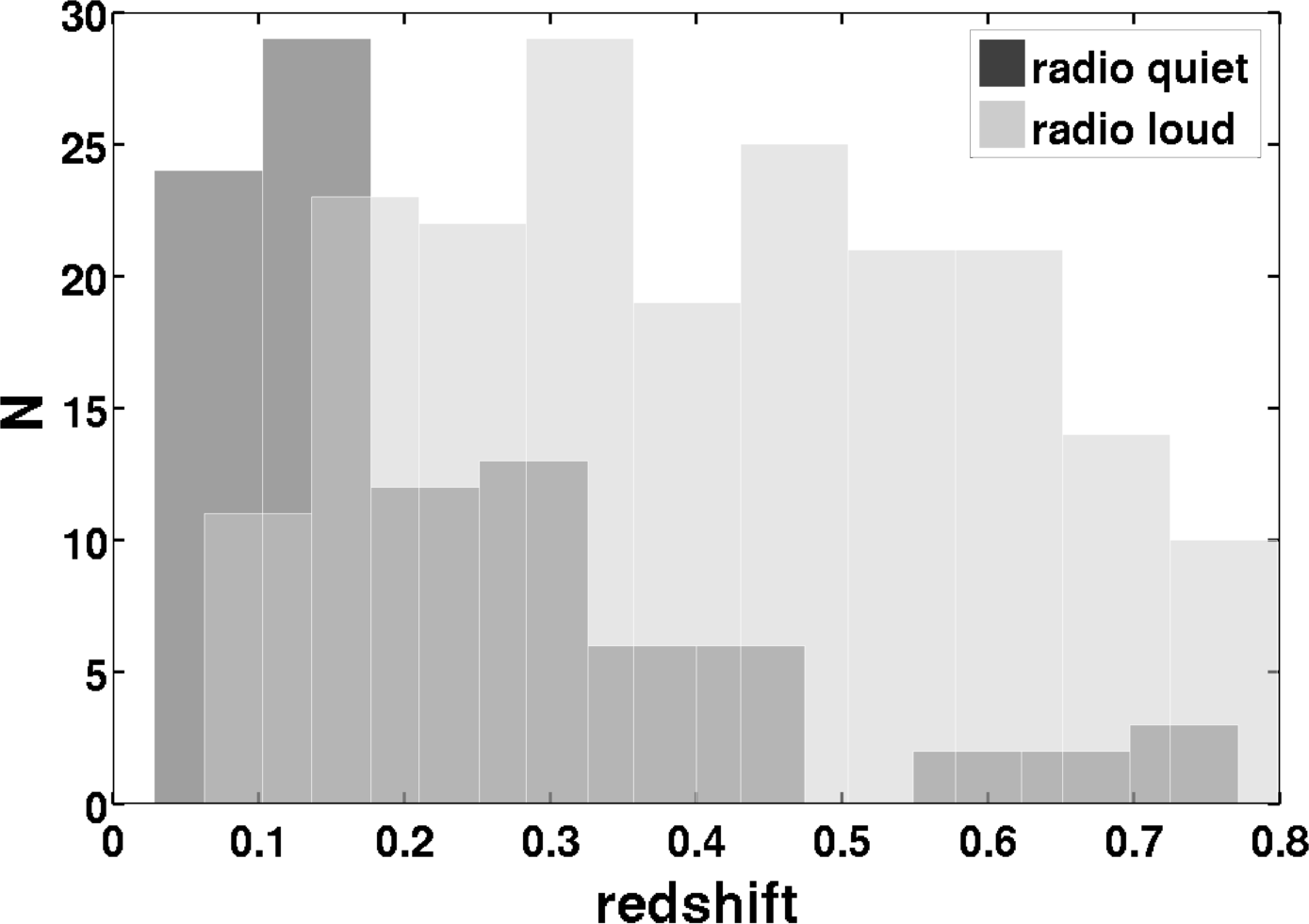}
  \caption{Redshift distributions of the radio-quiet and radio-loud subsamples.}
  \label{fig:rlrqhist}
\end{figure}

Mean redshifts and standard deviations for the different subsamples are given in Table~\ref{tab:subsamples}.
The K-S test for redshift suggests that our radio-quiet and radio-loud subsamples are not
originally from the same distribution (p=1.42$\times10^{-4}$). The distributions are shown in Figure~\ref{fig:rlrqhist}. 
In our sample, radio-quiet sources tend to lie closer than radio-loud sources.

The K-S test also suggests that radio-quiet and radio-loud subsamples are not drawn from the same luminosity distribution.
P-values for the K-S test are small, from 0.013 to $\sim10^{-26}$. Radio-loud sources are on average more 
luminous than radio-quiet sources. This holds for all wavebands. Examples of radio-loud and radio-quiet luminosity distributions are 
shown in Online Figure~\ref{fig:lumrlrqhist}. 

\onlfig{
\begin{figure*}[ht!]
\centering

  \subfloat{\includegraphics[width=0.505\textwidth]{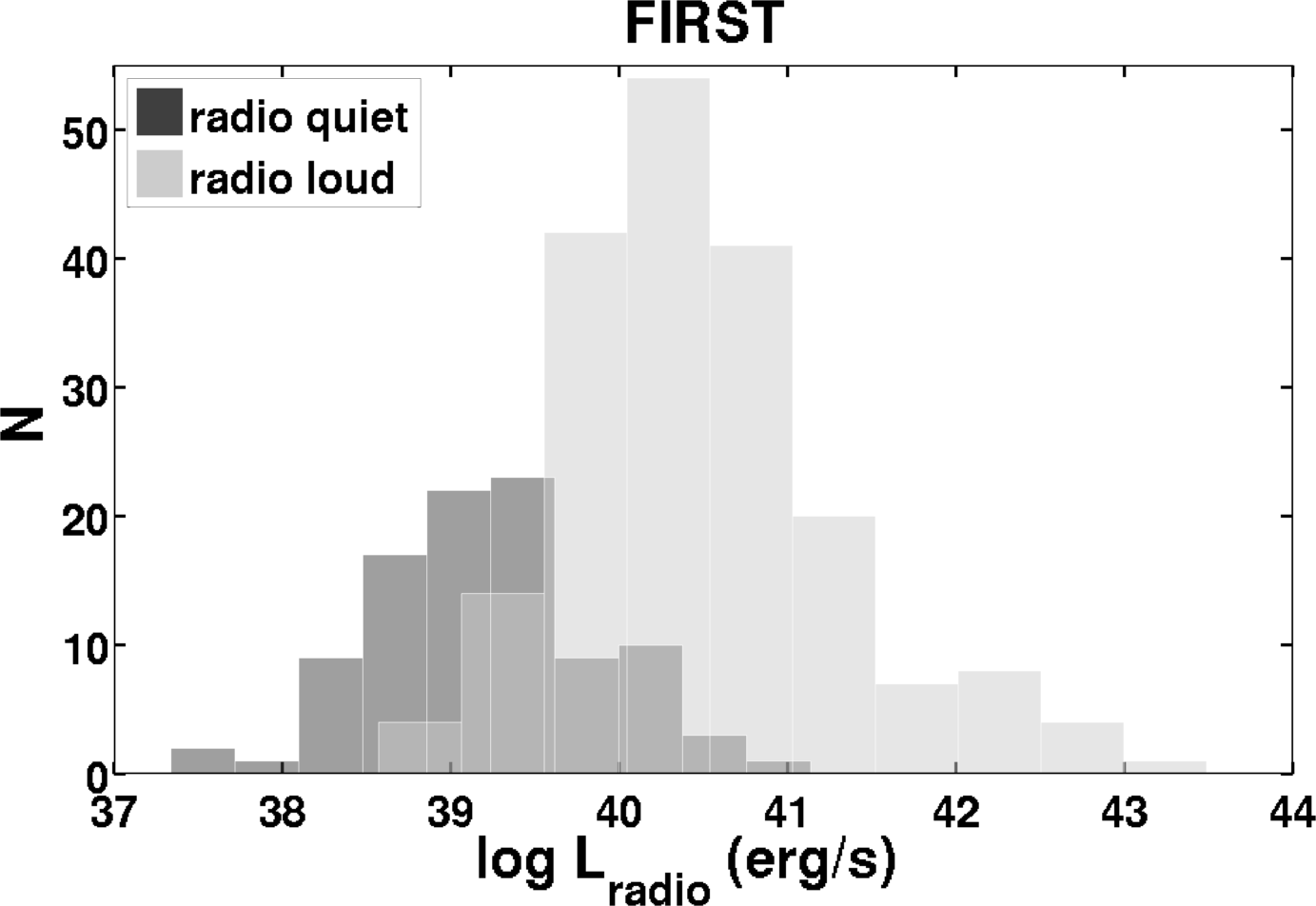}}
  \subfloat{\includegraphics[width=0.495\textwidth]{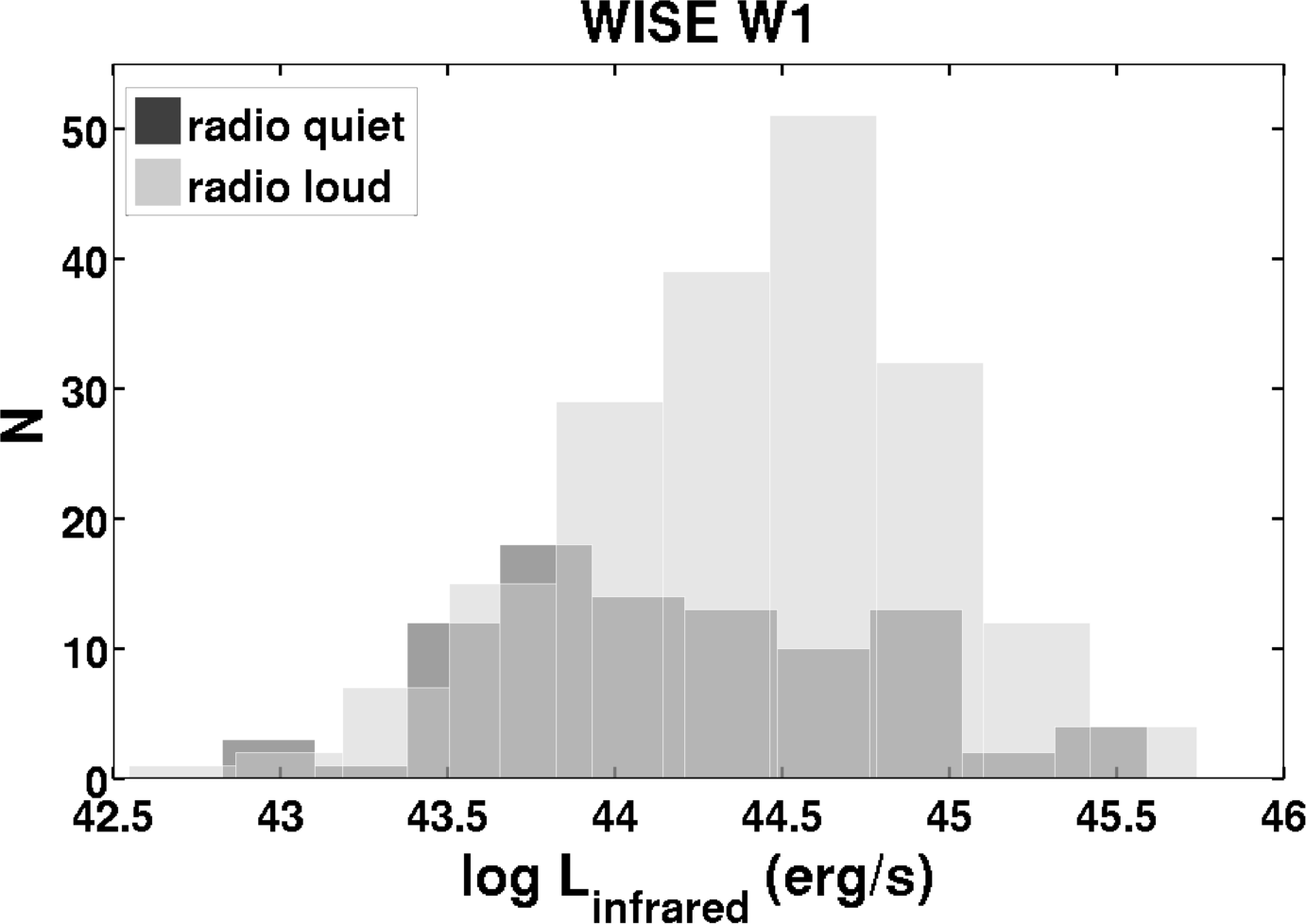}}

  \subfloat{\includegraphics[width=0.50\textwidth]{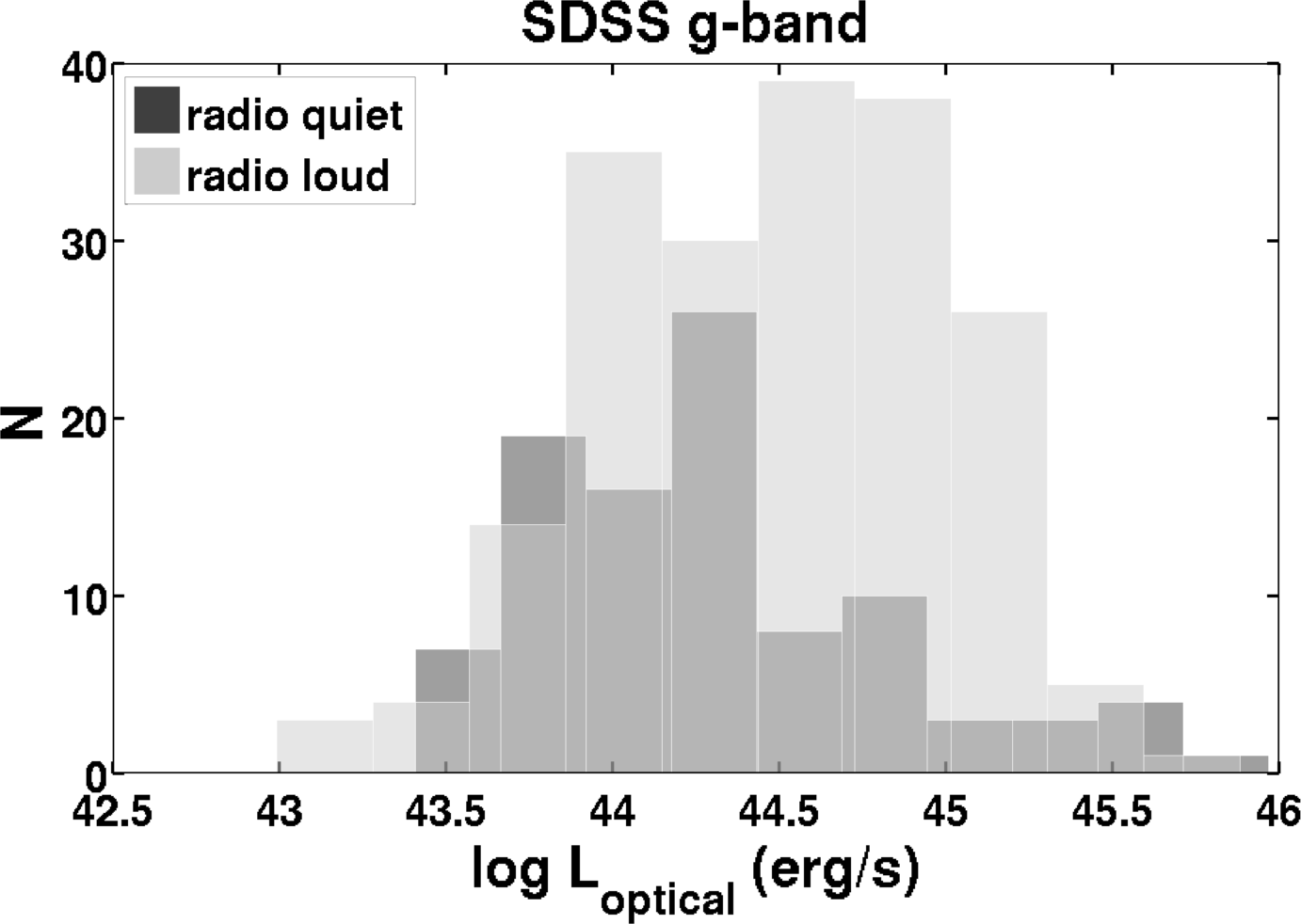}}
  \subfloat{\includegraphics[width=0.50\textwidth]{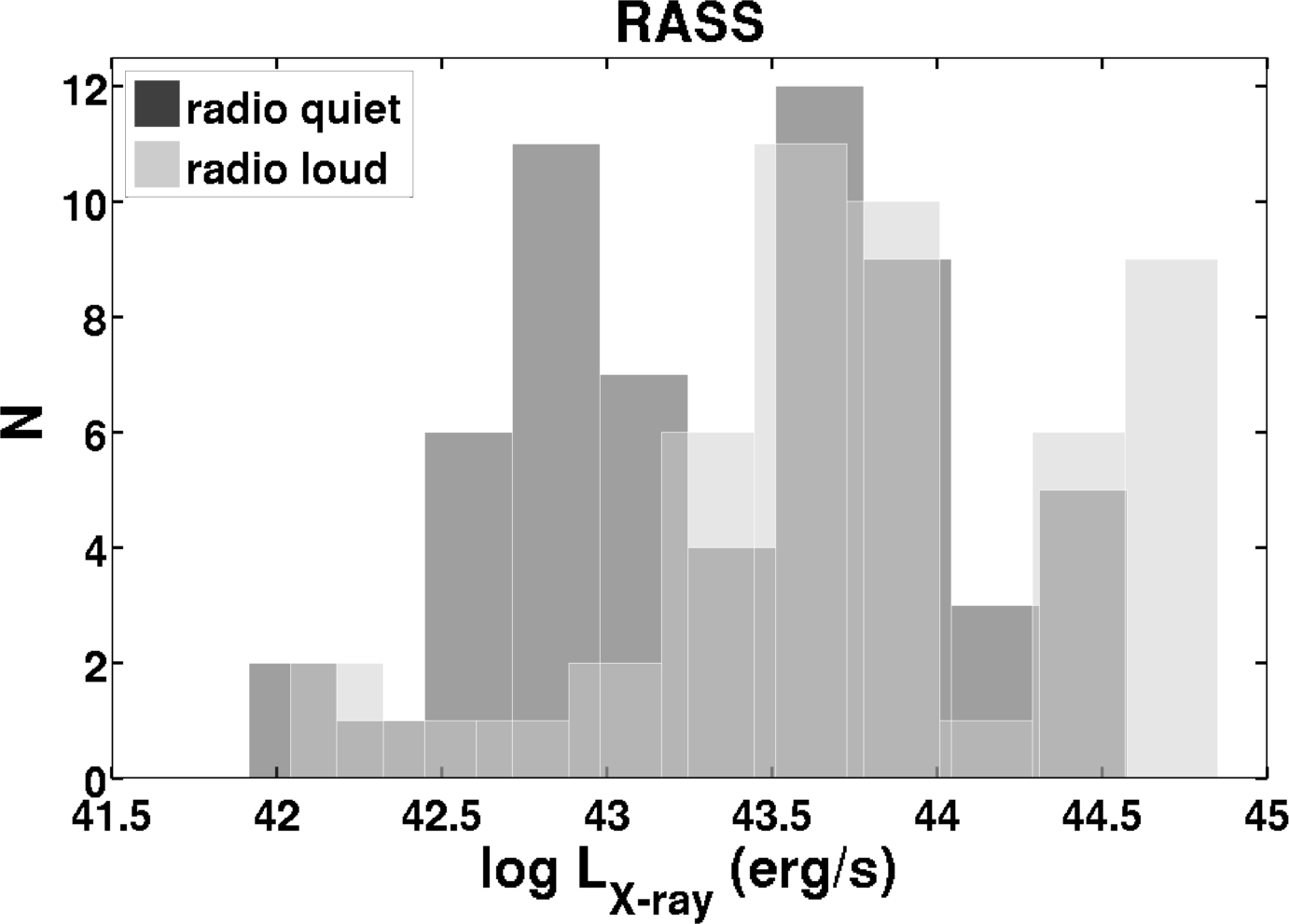}}

  \caption{Luminosity distributions of the radio-quiet and radio-loud subsamples for the wavebands for which we present 
  the correlation results; FIRST, WISE W1-band, SDSS $g$-band and RASS. }

\label{fig:lumrlrqhist}
\end{figure*}
}

\subsubsection{Black hole mass estimation}
\label{sec:bhmass}

There are three methods commonly used to estimate the black hole mass ($M_{\text{BH}}$) in AGN. These methods make use of the stellar velocity dispersion of 
the bulges ($\sigma_{\ast}$), the mass of the bulge ($M_{\text{bulge}}$, \citet{2009bentz1}), and FWHM(H$\beta$) or FWHM(H$\alpha$). We calculated $M_{\text{BH}}$ estimations 
using the FWHM(H$\beta$) -- luminosity mass scaling relation (see \citet{2005greene1} for more details)

\begin{equation} M_{\text{BH}} = (4.4 \pm 0.2) \times 10^6 \bigg( \frac{L_{5100}}{10^{44} \text{ ergs s}^{-1}} \bigg)^{0.64 \pm 0.02} \bigg( \frac{\text{FWHM}(H\beta)}{10^3 \text{ km s}^{-1}} \bigg)^2 M_{\sun}
\label{eq:mbh} \end{equation}

where $L_{\text{5100}}$ is the monochromatic luminosity at 5100\AA{}. $L_{\text{5100}}$ and FWHM(H$\beta$) values 
were taken from \citet{2006zhou1}. We used this method because there is no comprehensive enough information about $\sigma_{\ast}$ or 
$L_{\text{bulge}}$ for our sources. However, it does not take into account possible inclination effects caused by the geometry of the broad-line region (BLR) of the source and the viewing angle \citep{2011decarli2}. NLS1 sources also tend to lie below the normal $M_{\text{BH}} - \sigma_{\ast}$ and 
$M_{\text{BH}} - L_{\text{bulge}}$ relations \citep{2001mathur1, 2001laor1}.

We were able to estimate the $M_{\text{BH}}$ for 275 sources. The mean values and standard deviations are shown in 
Table~\ref{tab:subsamples}, and the black hole mass estimates for individual sources are listed in Online Table~\ref{tab:bigtable1}.

\begin{figure}[ht!]
 \includegraphics[width=0.49\textwidth]{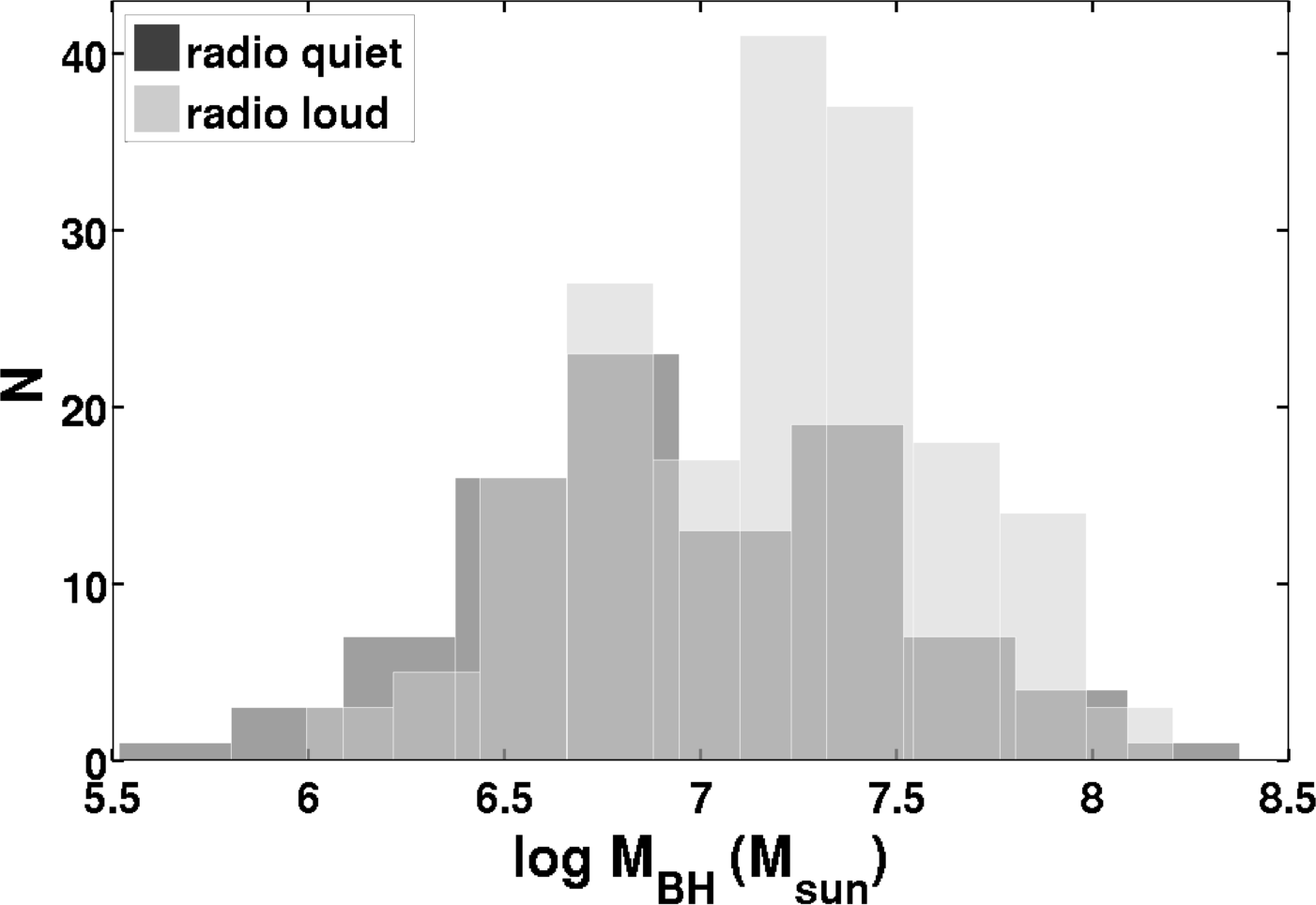}
\caption{Black hole mass distribution of the radio-quiet and radio-loud subsamples.}
 \label{fig:mbhdist}
\end{figure}

The K-S test for $M_{\text{BH}}$ suggests that the radio-quiet and radio-loud subsamples are not drawn from the same distribution (p=2.04$\times10^{-4}$). 
On average radio-loud sources have more massive black holes than radio-quiet sources. \citet{2004mclure1} got similar
results for a large sample of quasars (6099 radio-quiet and 436 radio-loud sources).
The $M_{\text{BH}}$ distributions of the radio-loud and radio-quiet subsamples are shown in Figure~\ref{fig:mbhdist}.

\subsection{Multiwavelength correlations}
\label{sec:mwcorr}

Results of multiwavelength correlations in literature are diverse and especially for NLS1 sources also very sparse.
\citet{2010arshakian1} found $L_{\text{R, total}}$ -- $L_{\text{O}}$ correlation for a sample of $\sim$100 quasars, and $L_{\text{R, jet}}$ -- $L_{\text{O}}$ 
correlation for BL Lac objects ($\sim$18). \citet{2012ballo1} concluded that in a sample of 852 quasars and 14 Seyfert galaxies 
there is $L_{\text{R}}$ -- $L_{\text{X}}$ correlation but no $L_{\text{O}}$ -- $L_{\text{X}}$ correlation, whereas \citet{2000brinkmann1} found a correlation
for both in a mixed sample. $L_{\text{R}}$ -- $L_{\text{X}}$ correlation has also been found in several other studies (\citet{2012younes1} for
six LINER 1 sources, \citet{2007panessa1} for 47 Seyfert galaxies and 33 low-luminosity radio galaxies, \citet{2009bianchi1} for 156 radio-quiet X-ray
unobscured AGN). Connection between radio and gamma-ray emission in radio-loud AGN has been confirmed by, for example, \citet{2001jorstad1}, \citet{2003lahteenmaki1}
and \citet{2011nieppola1}, and between mm and submm and gamma-ray emission by \citet{2012leontavares1}.

We used Pearson product-moment correlation coefficient (Pearson's r), Spearman's rank correlation coefficient (Spearman's $\rho$) and Pearson and 
Spearman partial correlation coefficients to study how the different wavebands are connected. The correlation is significant and the null hypothesis 
rejected if p$<$0.05.

We computed correlations for the flux densities for the whole sample and for all the subsamples individually. The correlations are shown in Table~\ref{tab:corrs} and the 
most significant results are shown in Online Table~\ref{tab:nor-compressed}. The flux densities used to calculate the results presented in 
Table~\ref{tab:nor-compressed} are the optical and radio flux densities used to calculate the radio-loudness, WISE W1 infrared flux density, and 
RASS X-ray flux density. We used SDSS {\it g}-band because it is closest to B-band which we used to calculate the radio-loudness. 
All the infrared and optical bands had very similar correlation results, therefore we present correlations for just one 
infrared and one optical band. 

The same correlations were computed also for luminosities, and in addition we determined the Pearson's r and Spearman's $\rho$ partial correlation coefficients for 
all the correlations shown in Table~\ref{tab:corrs}. For luminosities the used wavebands are FIRST, SDSS {\it g}-band,
WISE W1-band, and RASS. The results are again presented for one infrared and one optical band due
to the similarity of the results in all bands. For partial correlations we used redshift as the third parameter
to rule out the possibility that the possible correlation is not real but caused by redshift. The main results for partial correlations are shown in
Table~\ref{tab:par-compressed} and L--L plots in Online Figure~\ref{fig:llplots}. 
All other calculated multiwavelength correlations, e.g., the other optical and infrared wavebands, all luminosity correlations, 
and correlations for the X-ray selected samples did not differ significantly from the results presented here so they are omitted. 

In addition to flux density and luminosity correlations we studied correlations $M_{\text{BH}}$ -- $RL$, $M_{\text{BH}}$ -- $L$ (Table~\ref{tab:mbh-lum}),
and FWHM(H$\beta$) -- $L$ (Table~\ref{tab:fwhm-lum}). They are discussed in more detail in Sections~\ref{sec:mbhrlcorr},~\ref{sec:mbhlcorr}, and ~\ref{sec:fwhmlcorr}.

In order to get more insight into whether the origin of the emission at certain wavebands is the same in our radio-quiet and radio-loud subsamples,
we computed linear fits (y = ax + b) for flux densities and luminosities between different wavebands by using the least squares method.
Totally distinct slopes would suggest a different origin. Results for the fits for flux densities are shown in Online Table~\ref{tab:linearfitFLUX} and 
for luminosities in Online Table~\ref{tab:linearfitLUM}.

The multiwavelength correlations and linear fits are discussed in the next section.

\onlfig{
\begin{figure*}[ht!]
\centering

  \subfloat{\includegraphics[width=0.48\textwidth]{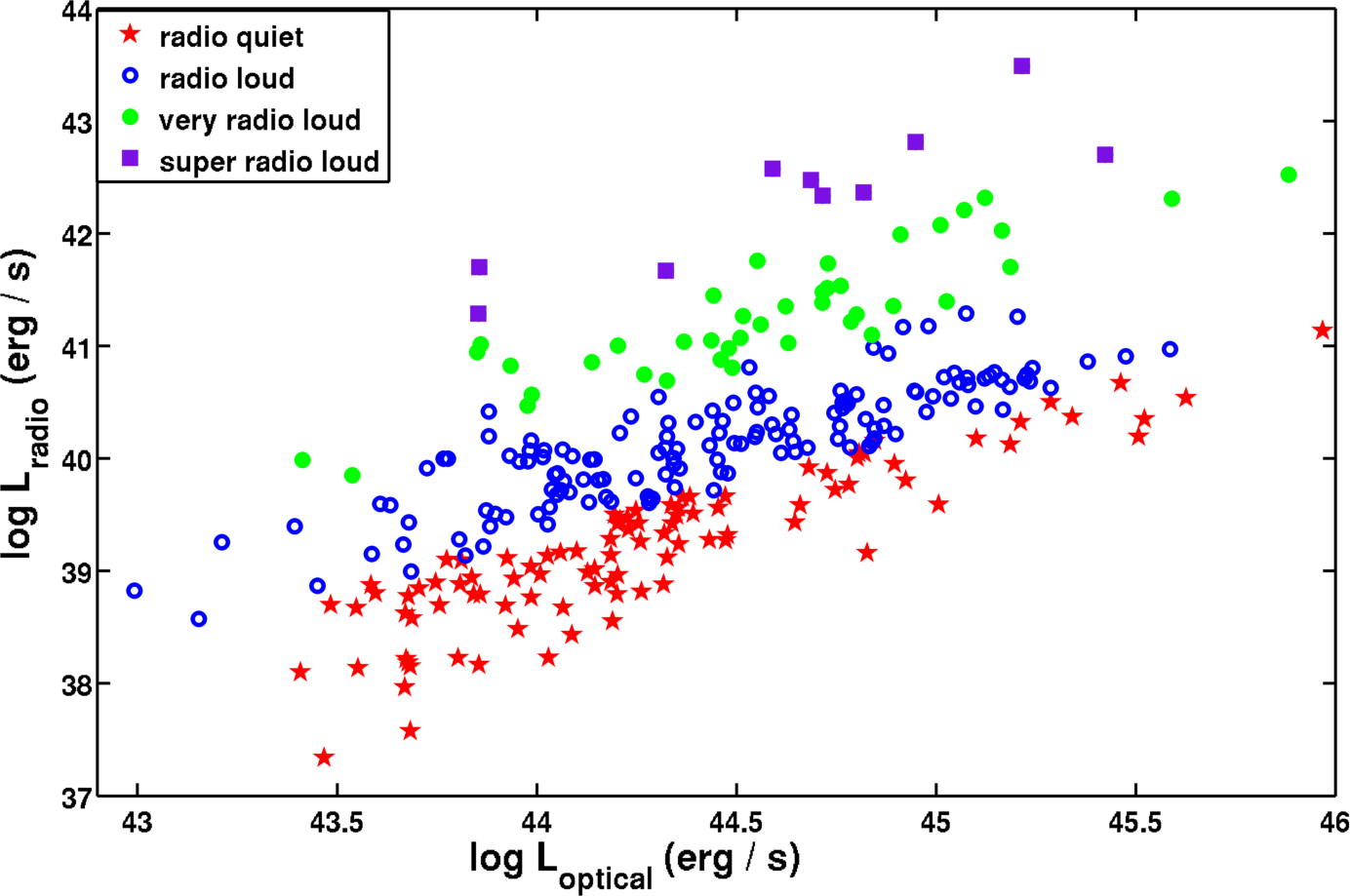}} \hspace{0.2cm}
  \subfloat{\includegraphics[width=0.49\textwidth]{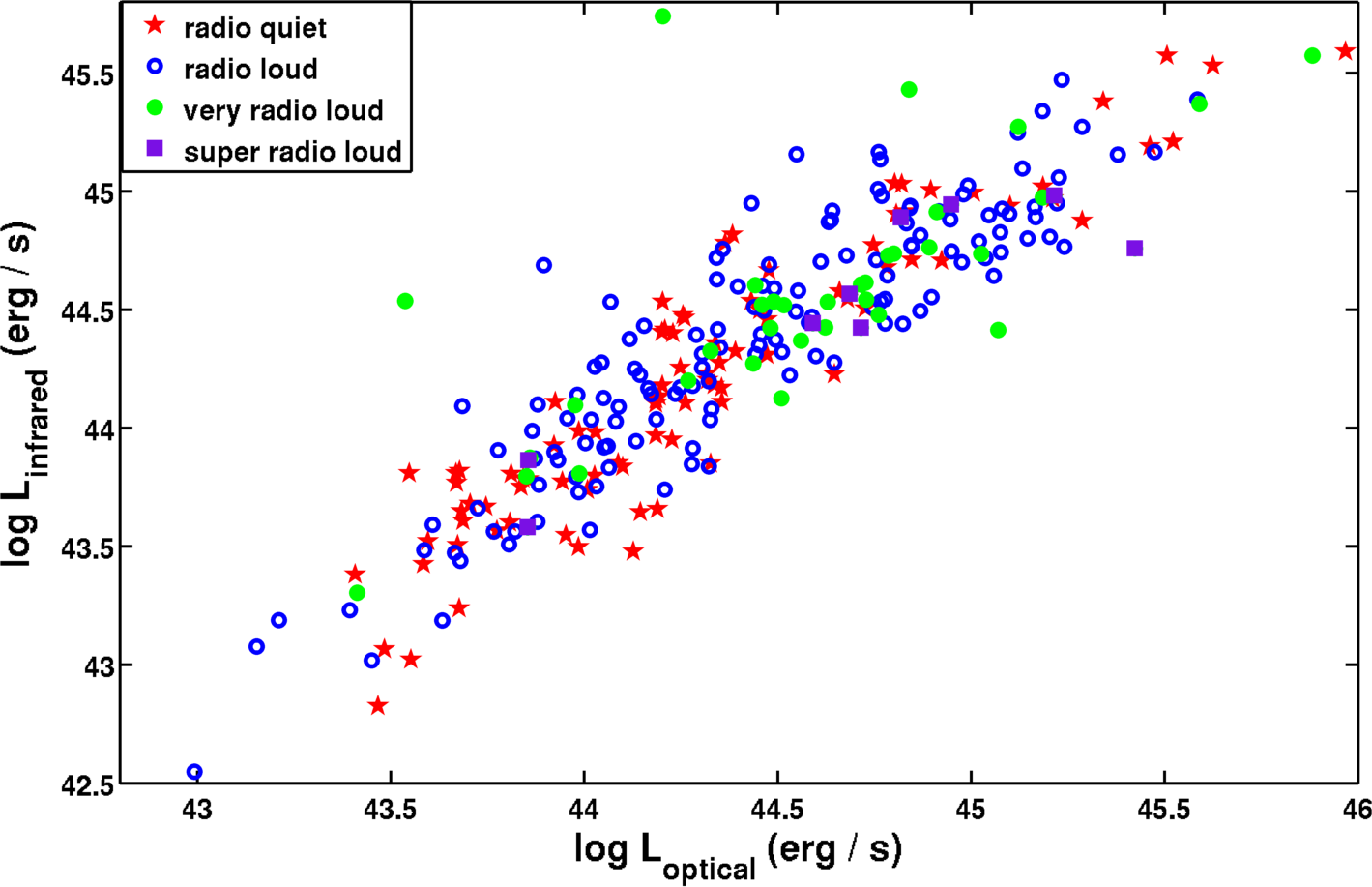}}
 
  \subfloat{\includegraphics[width=0.49\textwidth]{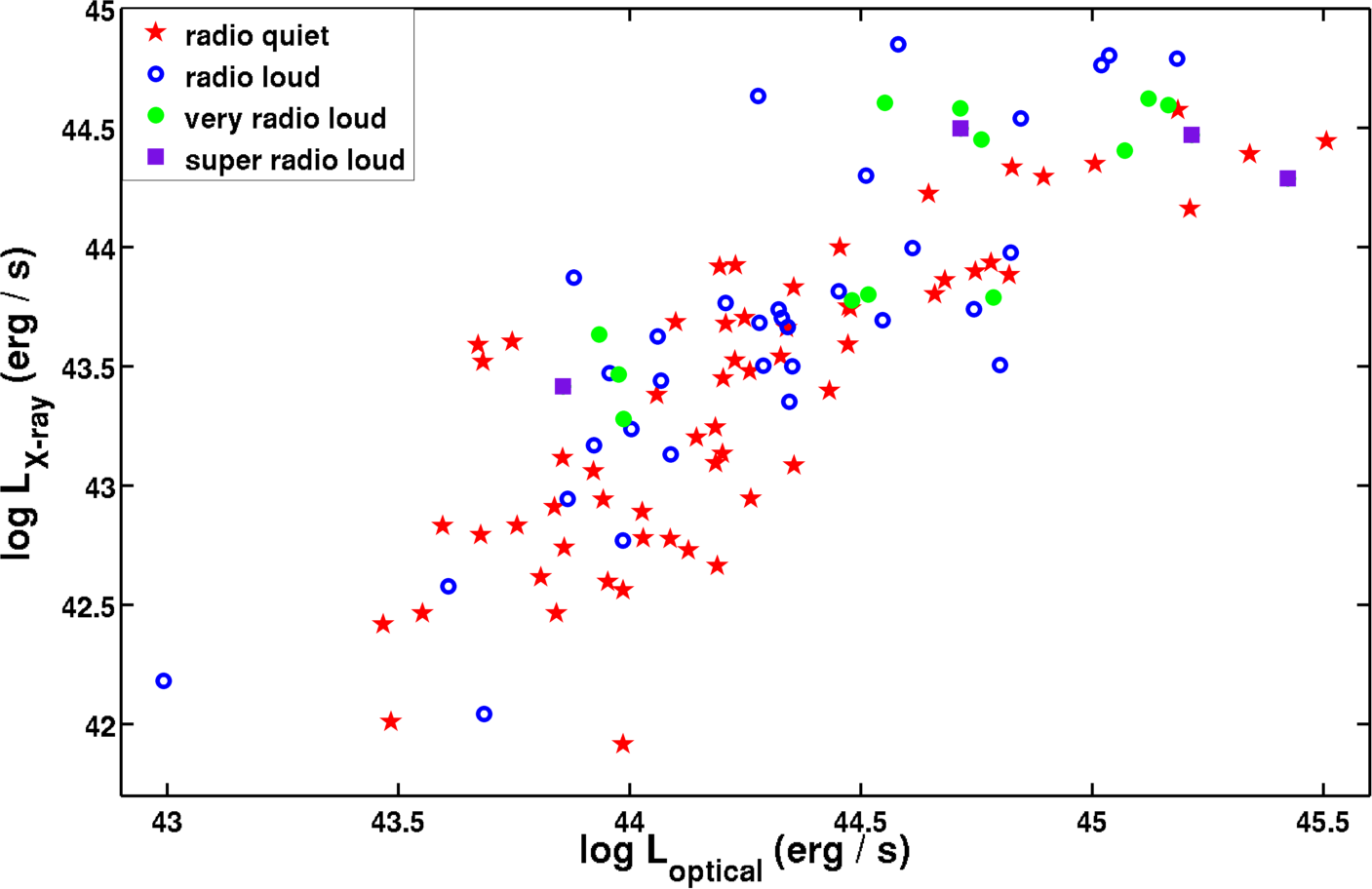}} \hspace{0.2cm}
  \subfloat{\includegraphics[width=0.49\textwidth]{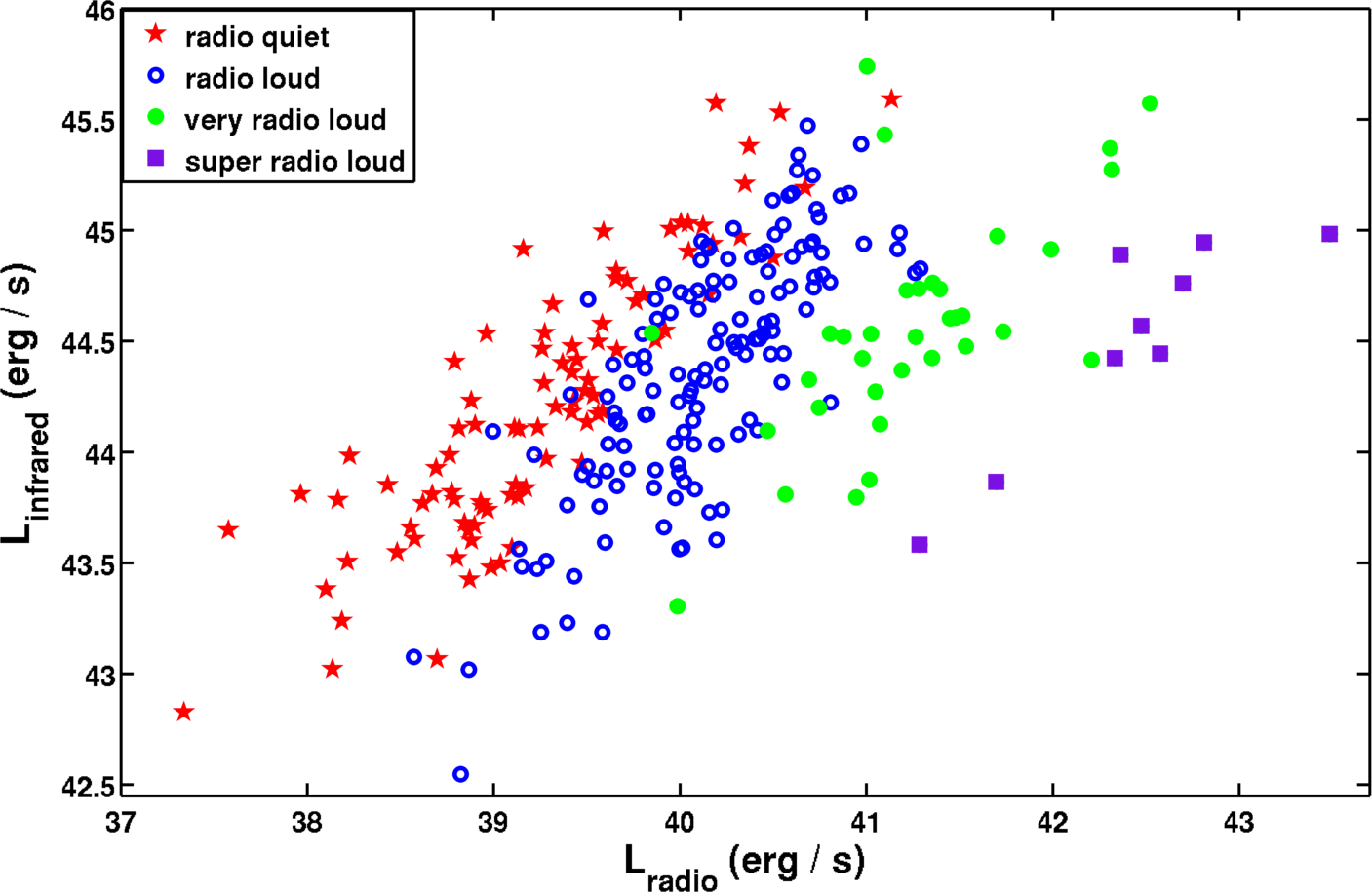}}
 
  \subfloat{\includegraphics[width=0.49\textwidth]{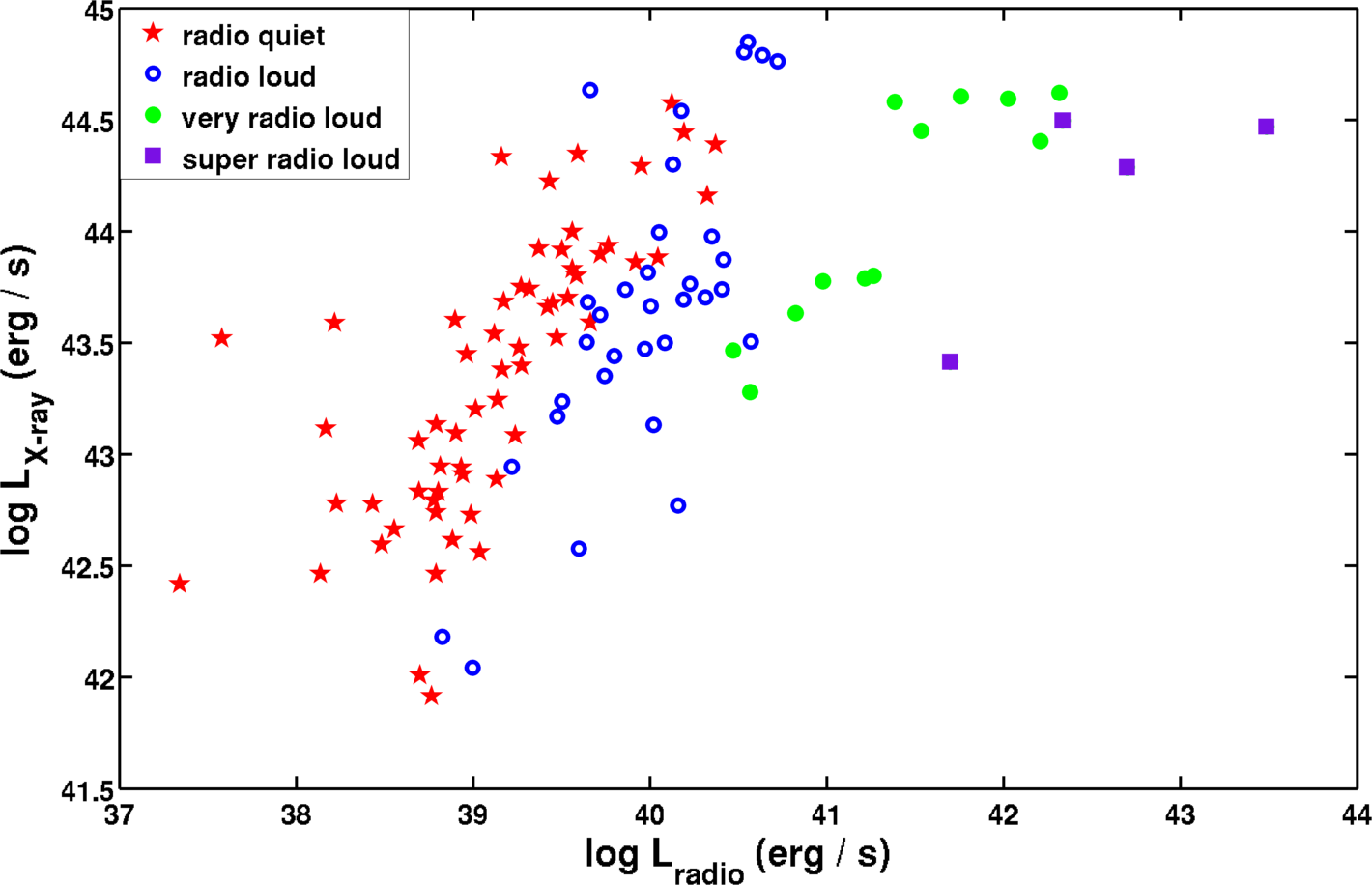}}

  \caption{Luminosity dependencies between the wavebands we used to compute the correlations. Subsamples are shown with different symbols and colors; radio-quiet: filled red stars, radio-loud: open blue
  circles, very radio-loud: filled green circles, and super radio-loud: filled purple squares.}
  
\label{fig:llplots}
\end{figure*}
}

\subsubsection{Radio-loud and radio-quiet sources}
\label{sec:rlrqmwcorr}

The results for flux density, luminosity, and partial luminosity correlations were similar.  

In the radio-loud subsample the optical band correlates rather well with the infrared (all subsamples) and X-ray (RL sample as a whole 
but not for VRL and SRL subsamples separately) bands. The correlation between the 
optical and radio bands becomes obvious only when looking at the VRL and SRL subsamples.
This suggests that in radio-loud sources radio, optical, and at least partially also X-ray emission come mostly from the jet as the jet emits low-energy 
synchrotron radiation (radio emission) and high-energy IC emission (X-ray emission). However, a portion of the X-ray emission
might also originate in the hot corona of the accretion disk via IC. Similarly, optical emission is supposedly composed of non-thermal emission from the jet and 
thermal emission from the accretion disk. Radio and infrared emission do not correlate (except for the SRL subsample which, however, is very small and therefore the correlation
cannot be considered convincing) which might indicate that infrared emission is rather of thermal than non-thermal origin,
for example, from the dusty torus (reprocessed emission from the accretion disk) or from star formation (reradiated starlight) \citep{2006bressan1}. 
Radio emission does not correlate with X-rays either. This is probably due to the very high and rapid X-ray variability exhibited by 
NLS1 sources (e.g. \citet{1999leighly1, 2000komossa1}). X-ray emission is variable in very short time scales, and since we have only one X-ray 
waveband observation per source, the observed X-ray flux value is highly coincidental. We also do not have any simultaneous radio and X-ray data.

In the radio-quiet subsample the optical band correlates very well with the infrared, and moderately well with radio and X-ray bands. Due to the radio-quietness of these 
sources it is likely that they do not host a jet or the jet is not powerful enough to dominate the emission. The predominant source 
of radio emission in the radio-quiet subsample could therefore be supernova remnants \citep{1992condon1}. In fact, radio emission does not correlate too well with the X-ray 
emission which probably is produced in the hot corona of the accretion disk via IC. As for the optical emission, it probably comes from the accretion disk. 
Radio and infrared bands correlate moderately well which suggests that infrared emission is generated by star formation as reradiated starlight \citep{2012botticella1}. 
However, dust heated by the AGN should also be taken into account when modeling both radio-quiet and radio-loud AGN \citep{2013mason1}.

Linear fits for flux densities and luminosities show very different slopes for the radio-quiet and radio-loud subsamples.
For example, $S_{\text{IR}}$ -- $S_{\text{R}}$ slope for the radio-quiet subsample is 0.759, whereas for the radio-loud
it is 0.070. Corresponding slopes for luminosities are 0.754 and 0.373. This suggests that the infrared
emission mechanism is different for the radio-quiet and radio-loud sources.

\begin{table}[ht]

\caption{Flux density and luminosity correlations between the different wavebands.}
\centering
\begin{tabular}{l l l l}
\hline\hline
\multicolumn{1}{c}{}                         & FIRST                                        & SDSS {\it g}               & RASS \\ \hline

FIRST                                        & -                                            & r, $\rho$, r$_{\text{x}}$, $\rho_{\text{x}}$ & r$_{\text{x}}$, $\rho_{\text{x}}$ \\ 
SDSS {\it u}/{\it g}/{\it r}/{\it i}/{\it z} & r\tablefootmark{a}, $\rho$\tablefootmark{b}, r$_{\text{x}}$\tablefootmark{c}, $\rho_{\text{x}}$\tablefootmark{d} & - & r$_{\text{x}}$, $\rho_{\text{x}}$ \\
WISE W1--4                             & r, $\rho$, r$_x$, $\rho_x$                   & r, $\rho$, r$_{\text{x}}$, $\rho_{\text{x}}$ & r$_{\text{x}}$, $\rho_{\text{x}}$ \\ 
RASS                                         & r$_{\text{x}}$, $\rho_{\text{x}}$            & r$_{\text{x}}$, $\rho_{\text{x}}$            & - \\ \hline

\end{tabular}
\label{tab:corrs}

\tablefoot{
\tablefoottext{a}{Pearson's r}
\tablefoottext{b}{Spearman's $\rho$}
\tablefoottext{c}{Pearson's r for X-ray sample}
\tablefoottext{d}{Spearman's $\rho$ for X-ray sample}
}

\end{table}

\onltab{
\begin{table*}[ht!]
\caption{Pearson's r and Spearman's $\rho$ flux density correlations and their p-values for the whole sample and the subsamples. Correlations in boldface have p$<$0.05}
\begin{tabular}{l l l l l l l}
\hline\hline
\multicolumn{1}{c}{}  & \multicolumn{2}{c}{$\log F_\text{O}$ - $\log F_\text{R}$}              & \multicolumn{2}{c}{$\log F_\text{O}$ - $\log F_{\text{IR}}$} &  \multicolumn{2}{c}{$\log F_\text{O}$ - $\log F_\text{X}$}                                                  \\          
sample                & Pearson's r (p)                          & Spearman's $\rho$  (p)      & r   (p)                       & $\rho$ (p)                   & r (p)                        & $\rho$ (p)                     \\   \hline 
all                   & -0.024 (0.678)                           & 0.053 (0.363)               & {\bf 0.810} ($\sim10^{-67}$)  & {\bf 0.787} ($\sim10^{-61}$) & {\bf 0.556} ($\sim10^{-10}$) & {\bf 0.548} ($\sim10^{-10}$)   \\ 
RQ                    & {\bf 0.645} ($\sim10^{-13}$)             & {\bf 0.658} ($\sim$0)       & {\bf 0.840} ($\sim10^{-25}$)  & {\bf 0.843} ($\sim10^{-25}$) & {\bf 0.592} ($\sim10^{-7}$)  & {\bf 0.597} ($\sim10^{-7}$)    \\ 
RL                    & {\bf 0.248} ($\sim10^{-4}$)              & {\bf 0.304} ($\sim10^{-5}$) & {\bf 0.630} ($\sim10^{-22}$)  & {\bf 0.601} ($\sim10^{-20}$) & {\bf 0.456} (0.001)          & {\bf 0.484} ($\sim10^{-4}$)    \\ 
VRL                   & {\bf 0.560} ($\sim10^{-5}$)              & {\bf 0.609} ($\sim10^{-6}$) & {\bf 0.509} ($\sim10^{-4}$)   & {\bf 0.581} ($\sim10^{-6}$)  & {\bf 0.627} (0.009)          & {\bf 0.579} (0.021)            \\ 
SRL                   & {\bf 0.744} (0.014)                      & 0.467  (0.178)              & {\bf 0.918} ($\sim10^{-4}$)   & {\bf 0.855} (0.004)          & 0.115  (0.885)               & -0.200 (0.917)                 \\ \hline

\end{tabular}
\label{tab:nor-compressed}
\end{table*} 

\begin{table*}[ht!]
\begin{tabular}{l l l l l}
\hline\hline
\multicolumn{1}{c}{}  & \multicolumn{2}{c}{$\log F_\text{R}$ - $\log F_{\text{IR}}$}              &  \multicolumn{2}{c}{$\log F_\text{R}$ - $\log F_\text{X}$}  \\         
sample                & Pearson's r (p)             & Spearman's $\rho$  (p)      & r (p)           & $\rho$ (p)          \\ \hline   
all                   & -0.055 (0.360)              & -0.017 (0.782)              & 0.047 (0.624)   &  0.130 (0.179)      \\ 
RQ                    & {\bf 0.548} ($\sim10^{-8}$)  & {\bf 0.571} ($\sim10^{-9}$)  & 0.251 (0.053)   & {\bf 0.333} (0.010) \\ 
RL                    &  0.131 (0.071)              & {\bf 0.142} (0.049)         & 0.213 (0.141)   &  0.255 (0.077)      \\ 
VRL                   & 0.181 (0.202)               & {\bf 0.289} (0.039)         & 0.229 (0.394)   &  0.159 (0.556)      \\ 
SRL                   & {\bf 0.687} (0.028)         & 0.418 (0.232)               & -0.205 (0.795) &   0.200 (0.917)      \\ \hline

\end{tabular}
\end{table*}

}

\begin{table*}[ht!]
\caption{Pearson's r and Spearman's $\rho$ partial luminosity correlations and their p-values for the whole sample and the subsamples. Correlations in boldface have p$<$0.05}
\begin{tabular}{l l l l l l l} 
\hline\hline
\multicolumn{1}{c}{}  & \multicolumn{2}{c}{$\log L_\text{O}$ - $\log L_\text{R}$}                 & \multicolumn{2}{c}{$\log L_\text{O}$ - $\log L_{\text{IR}}$}              &  \multicolumn{2}{c}{$\log L_\text{O}$ - $\log L_\text{X}$}                \\           
sample                & Pearson's r (p)             & Spearman's $\rho$  (p)      & r (p)                        & $\rho$  (p)                  & r (p)                       & $\rho$ (p)                  \\   \hline
all                   & 0.033 (0.571)               & -0.044 (0.457)              & {\bf 0.701} ($\sim10^{-43}$) & {\bf 0.699} ($\sim10^{-42}$) & {\bf 0.484} ($\sim10^{-7}$) & {\bf 0.456} ($\sim10^{-7}$) \\ 
RQ                    & {\bf 0.507} ($\sim10^{-7}$) & {\bf 0.528} ($\sim10^{-8}$) & {\bf 0.767} ($\sim10^{-18}$) & {\bf 0.734} ($\sim10^{-16}$) & {\bf 0.476} ($\sim10^{-4}$) & {\bf 0.384} (0.003)         \\ 
RL                    & {\bf 0.223} (0.002)         & {\bf 0.160} (0.026)         & {\bf 0.607} ($\sim10^{-20}$) & {\bf 0.575} ($\sim10^{-18}$) & {\bf 0.538} ($\sim10^{-5}$) & {\bf 0.451} (0.001)         \\ 
VRL                   & {\bf 0.527} ($\sim10^{-5}$) & {\bf 0.577}($\sim10^{-5}$)  & {\bf 0.448} (0.001)          & {\bf 0.605} ($\sim10^{-6}$)  & 0.422 (0.117)               & 0.233 (0.402)               \\
SRL                   & {\bf 0.744} (0.021)         & {\bf 0.699} (0.036)         & {\bf 0.950} ($\sim10^{-5}$)  & {\bf 0.931} ($\sim10^{-4}$)  & 0.505 (0.663)               & -0.218 (0.860)              \\ \hline

\end{tabular}
\label{tab:par-compressed}
\end{table*} 

\begin{table*}[ht!]
\begin{tabular}{l l l l l} 
\hline\hline
\multicolumn{1}{c}{} & \multicolumn{2}{c}{$\log L_\text{R}$ - $\log L_{\text{IR}}$}                &  \multicolumn{2}{c}{$\log L_\text{R}$ - $\log L_\text{X}$} \\           
sample                & Pearson's r (p)              & Spearman's $\rho$  (p)      & r (p)          & $\rho$  (p)         \\ \hline
all                   & 0.030   (0.613)              & -0.078  (0.192)             & 0.151 (0.119)  & 0.184 (0.057)       \\ 
RQ                    & {\bf 0.546} ($\sim10^{-8}$)  & {\bf 0.380} ($\sim10^{-4}$) & 0.230 (0.079)  & {\bf 0.314} (0.016) \\
RL                    & 0.101 (0.163)                & 0.018 (0.800)               & 0.201 (0.171)  & 0.277 (0.057)       \\ 
VRL                   & 0.094 (0.518)                & 0.269 (0.059)               & 0.218 (0.436)  & 0.310 (0.261)       \\ 
SRL                   & {\bf 0.676} (0.048)          & {\bf 0.806} (0.009)         & 0.171 (0.890)  & N/A (N/A)           \\ \hline

\end{tabular}
\end{table*}

\onltab{
\begin{table*}[ht]
\begin{minipage}{1\textwidth}
\caption{Linear fits of the flux densities for the radio-quiet and radio-loud subsamples.}
\centering
\begin{tabular}{l}

\hline\hline
\multicolumn{1}{l}{RADIO QUIET SAMPLE}\\ \hline
$\log S_{\text{O}}$ = 0.994($\pm$0.012)$\log S_{\text{R}}$ - 0.604($\pm$0.034) \\ 
$\log S_{\text{IR}}$  = 0.759($\pm$0.013)$\log S_{\text{R}}$ - 0.447($\pm$0.039) \\ 
$\log S_{\text{X}}$   = 0.344($\pm$0.018)$\log S_{\text{R}}$ - 5.808($\pm$0.051) \\ 
$\log S_{\text{IR}}$  = 0.753($\pm$0.013)$\log S_{\text{O}}$   - 0.013($\pm$0.047) \\ 
$\log S_{\text{X}}$   = 0.568($\pm$0.015)$\log S_{\text{O}}$   - 4.863($\pm$0.051) \\
$\log S_{\text{X}}$   = 0.648($\pm$0.018)$\log S_{\text{IR}}$    - 5.119($\pm$0.046) \\ \hline
\multicolumn{1}{l}{}\\ \hline

\multicolumn{1}{l}{RADIO LOUD SAMPLE}\\ \hline
$\log S_{\text{O}}$ = 0.132($\pm$0.002)$\log S_{\text{R}}$ - 3.820($\pm$0.005) \\ 
$\log S_{\text{IR}}$  = 0.070($\pm$0.002)$\log S_{\text{R}}$ - 2.984($\pm$0.005) \\ 
$\log S_{\text{X}}$   = 0.119($\pm$0.010)$\log S_{\text{R}}$ - 6.729($\pm$0.023) \\ 
$\log S_{\text{IR}}$  = 0.638($\pm$0.004)$\log S_{\text{O}}$   - 0.513($\pm$0.018) \\ 
$\log S_{\text{X}}$   = 0.546($\pm$0.024)$\log S_{\text{OP}}$   - 4.836($\pm$0.094) \\ 
$\log S_{\text{X}}$   = 0.311($\pm$0.019)$\log S_{\text{IR}}$    - 6.046($\pm$0.057) \\ \hline

\end{tabular}
\label{tab:linearfitFLUX}
\end{minipage}
\end{table*} 
}

\onltab{
\begin{table*}[ht]
\begin{minipage}{1\textwidth}
\caption{Linear fits of the luminosities for the radio-quiet and radio-loud subsamples.}
\centering
\begin{tabular}{l}

\hline\hline

\multicolumn{1}{l}{RADIO QUIET SAMPLE}\\ \hline
$\log L_{\text{O}}$ = 0.705($\pm$0.006)$\log L_{\text{R}}$ + 16.654($\pm$0.249) \\
$\log L_{\text{IR}}$  = 0.754($\pm$0.006)$\log L_{\text{R}}$ + 14.629($\pm$0.226) \\ 
$\log L_{\text{X}}$   = 0.754($\pm$0.008)$\log L_{\text{R}}$ + 13.882($\pm$0.328) \\ 
$\log L_{\text{IR}}$  = 1.012($\pm$0.010)$\log L_{\text{O}}$   - 0.562($\pm$0.425) \\ 
$\log L_{\text{X}}$   = 1.107($\pm$0.013)$\log L_{\text{O}}$   - 5.606($\pm$0.572) \\ 
$\log L_{\text{X}}$   = 0.908($\pm$0.011)$\log L_{\text{IR}}$    + 3.308($\pm$0.504) \\ \hline
\multicolumn{1}{l}{}\\ \hline

\multicolumn{1}{l}{RADIO LOUD SAMPLE}\\ \hline
$\log L_{\text{O}}$ = 0.404($\pm$0.001)$\log L_{\text{R}}$ + 28.118($\pm$0.046) \\ 
$\log L_{\text{IR}}$  = 0.373($\pm$0.001)$\log L_{\text{R}}$ + 29.325($\pm$0.042) \\ 
$\log L_{\text{X}}$   = 0.419($\pm$0.006)$\log L_{\text{R}}$ + 26.810($\pm$0.235) \\ 
$\log L_{\text{IR}}$  = 0.915($\pm$0.003)$\log L_{\text{O}}$   + 3.710($\pm$0.126) \\ 
$\log L_{\text{X}}$   = 1.121($\pm$0.016)$\log L_{\text{O}}$   - 5.964($\pm$0.719) \\ 
$\log L_{\text{X}}$   = 0.924($\pm$0.012)$\log L_{\text{IR}}$    + 2.841($\pm$0.554) \\ \hline

\end{tabular}
\label{tab:linearfitLUM}
\end{minipage}
\end{table*} 

}

\subsection{Principal component and cluster analysis}
\label{sec:pca}

Principal component analysis (PCA) is a statistical method used to simplify large amounts of data. It converts a set of possibly correlated variables into a
set of uncorrelated variables called principal components, or eigenvectors. The first principal component accounts for as much of the variability in the data
as possible. The second principal component has as large a variance as possible while still being orthogonal to the first principal component, and so on. This
method makes it possible to find underlying connections and the most dominant variables in a data set, and possibly helps to identify the physical properties
connected with each eigenvector. A good overview of the PCA in astronomy can be found in \citet{1999francis1}.

\citet{1992boroson1} used PCA to study the optical properties of 87 quasi-stellar objects (QSO). In their study Eigenvector 1 (EV1) is dominated by the anticorrelation between
the strength of Fe II, and the strength of [OIII] $\lambda$5007 and FWHM(H$\beta$), and Eigenvector 2 (EV2) distinguishes between the strength of He II $\lambda$4686
and optical luminosity. This study was continued in \citet{2002boroson1} where 75 sources were added to the original sample. The results for the first
two principal components were similar to the earlier study. They suggest that EV1 corresponds closely to the Eddington ratio, $L/L_{\text{Edd}}$,
and EV2 to the accretion rate. \citet{2012xu1} studied a sample of narrow-line and broad-line Seyfert 1 galaxies using PCA. 
Their results were consistent with \citet{1992boroson1} and \citet{2002boroson1}. In a study of 110 soft X-ray -selected AGN, of which
about half were NLS1 galaxies, \citet{2004grupe1} found the EV1 to be similar to the EV1 in \citet{2002boroson1}. EV2 in their study correlated
strongly with the black hole mass.

We performed weighted principal component analysis (PCA) using the pca{\footnote{http://www.mathworks.se/help/stats/pca.html}} function in MATLAB Statistics Toolbox for the whole sample (97 sources), the radio-loud subsample (41 sources), the radio-quiet subsample (56 sources), and the corresponding modified samples 
(super radio-loud outlier, SDSS J104732.68+472532.1, removed). We carried out PCA with seven variables for all samples. The variables used were the radio flux density (FIRST), 
infrared flux density (WISE W1-band), optical flux density (SDSS $g$-band), X-ray flux density (RASS), $M_{\text{BH}}$, FWHM(H$\beta$) (H$\beta$ broad component FWHM), and 
R4570 (optical Fe II strength relative to broad H$\beta$ component). We included R4570 to the PCA because it is possibly related to the radio emission. For example,
\citet{2008yuan1} found that the optical Fe II emission is on average stronger for radio-loud NLS1 sources than for the NLS1 population in general. Values for FWHM(H$\beta$) and R4570 were 
taken from \citet{2006zhou1}. To start with, we tried several combinations of variables and, for example, correlated $M_{\text{BH}}$, in the cases when it 
was left out of the PCA, with the Eigenvectors along the lines of \citet{2012xu1} and \citet{2004grupe1}. Unfortunately, most of these experiments did not yield any convincing 
results so instead we changed the focus of the analysis to how the different properties are linked to each other, and therefore decided to include 
$M_{\text{BH}}$ in the PCA. This proved to be a more informative approach. 

Results only for the first and the second principal components are presented because the results for the subsequent components were mixed and no conclusions could
be drawn from them. The PCA coefficients are listed in Tables~\ref{tab:pca7ev1} and ~\ref{tab:pca7ev2}. In the tables, 
the coefficients have been grouped together based on their sign, i.e. whether they correlate or anticorrelate with the Eigenvector. In this way, it is easier
to see which properties might be linked to each other.

We also tried cluster analysis for our whole sample in order to see if there are any distinguishable groups within our sample. We tried both hierarchical clustering and
k-Means clustering with five variables: radio flux density (FIRST), infrared flux density (WISE W1-band), optical flux density (SDSS $g$-band), X-ray flux density (RASS), 
and $M_{\text{BH}}$. Cluster analysis did not yield any compelling results.

\begin{table}[ht]
\begin{minipage}{0.5\textwidth}
\caption{Results of the principal component analysis with seven variables, eigenvector 1. The coefficients have been grouped together based on their sign.}
\centering
\begin{tabular}{l l l}

\hline\hline

- & sample & + \\ \hline

IR -0.48          & All (35\%)  & radio 0.20               \\ 
optical -0.52     &       & $\log M_{\text{BH}}$ 0.45 \\
X-ray -0.39       &       & FWHM(H$\beta$) 0.32      \\
                  &       & R4570 0.04               \\
& & \\
IR -0.50         & All, modified (34\%)  & radio 0.10 \\
optical -0.55    &                 & $\log M_{\text{BH}}$ 0.44 \\
X-ray -0.41      &                 & FWHM(H$\beta$) 0.28 \\
                 &                 & R4570 0.06 \\
& & \\
IR -0.53          & Radio-loud (37\%) & radio 0.21 \\
optical -0.50     &            & $\log M_{\text{BH}}$ 0.49 \\
X-ray -0.16       &            & FWHM(H$\beta$) 0.39 \\ 
R4570 -0.08       &            &  \\
                  &            &  \\
& & \\
radio -0.01       & Radio-loud,  & $\log M_{\text{BH}}$ 0.48 \\
IR -0.57          & modified (35\%)      & FWHM(H$\beta$) 0.36 \\
optical -0.54     &                      &  \\
X-ray -0.15       &                      &  \\
R4570 -0.07       &                      &  \\
& & \\
radio -0.40       & Radio-quiet (37\%)& $\log M_{\text{BH}}$ 0.35 \\
IR -0.48          &             & FWHM(H$\beta$) 0.12 \\
optical -0.56     &             & R4570 0.19 \\
X-ray -0.35       &             &  \\ \hline

\end{tabular}
\label{tab:pca7ev1}
\end{minipage}
\end{table} 

\begin{table}[ht]
\begin{minipage}{0.5\textwidth}
\caption{Results of the principal component analysis with seven variables, eigenvector 2. The coefficients have been grouped together based on their sign.}
\centering
\begin{tabular}{l l l}

\hline\hline
- & sample & + \\ \hline
R4570 -0.52       & All (22\%) & radio 0.21              \\ 
                  &       & IR 0.31                   \\
                  &       & optical 0.28             \\
                  &       & X-ray 0.14                \\
                  &       & $\log M_{\text{BH}}$ 0.28 \\
                  &       & FWHM(H$\beta$) 0.64     \\
& & \\
radio -0.02      & All, modified (22\%)  & IR 0.27 \\
R4570 -0.54      &                 & optical 0.23  \\
                 &                 & X-ray 0.09 \\
                 &                 &  $\log M_{\text{BH}}$ 0.31 \\
                 &                 & FWHM(H$\beta$) 0.69 \\
& & \\
X-ray -0.21      & Radio-loud (20\%)& radio 0.29 \\
R4570 -0.56      &            & IR 0.29\\
                 &            & optical 0.41 \\ 
                 &            &  $\log M_{\text{BH}}$ 0.001 \\
                 &            & FWHM(H$\beta$) 0.55 \\
& & \\
X-ray -0.23               & Radio-loud,  & radio 0.04 \\
 $\log M_{\text{BH}}$ -0.06 & modified (19\%)      & IR 0.16 \\
R4570 -0.68               &                      & optical 0.32 \\
                          &                      & FWHM(H$\beta$) 0.59 \\
& & \\
X-ray -0.06         & Radio-quiet (24\%) & radio 0.11 \\
R4570 -0.41         &             & IR 0.25 \\  
                    &             & optical 0.06 \\
                    &             &  $\log M_{\text{BH}}$ 0.48 \\ 
                    &             &  FWHM(H$\beta$) 0.72 \\ \hline

\end{tabular}
\label{tab:pca7ev2}
\end{minipage}
\end{table} 

\subsubsection{Eigenvector 1}

EV1 accounts for 34\% -- 37\% of the variance. In all samples and subsamples EV1 distinguishes between $M_{\text{BH}}$ and 
both optical and infrared emissions. The latter two are always connected. X-ray emission behaves similarly to infrared and optical but contributes less. Radio 
emission is rather insignificant except in the radio-quiet subsample where it is connected with optical and infrared, confirming the suggestion that in 
the radio-quiet sources radio emission is rather of stellar origin. FWHM(H$\beta$) is strongly connected with $M_{\text{BH}}$ except in the radio-quiet subsample where it is less significant. 
R4570 is rather insignificant in all samples.

In all cases for EV1 the optical and infrared wavebands seem to be tightly connected, and opposite to $M_{\text{BH}}$ and FWHM(H$\beta$).
In the radio-quiet subsample also radio emission is connected with optical and infrared. In all cases the X-ray band behaves similarly to infrared and optical. These results suggest that optical
and infrared emission (and up to some extent X-ray emission) -- and in radio-quiet sources also the radio emission -- are either of thermal or stellar origin i.e. not generated in a jet.
The significance of $M_{\text{BH}}$ implies that EV1 might be similar to EV2 in \citet{2004grupe1}.

\subsubsection{Eigenvector 2}

19\%-24\% of the variance is explained by EV2. It clearly distinguishes
between R4570 and FWHM(H$\beta$) in all samples. To some extent also infrared and optical seem to be connected to FWHM(H$\beta$).
Since EV2 is dominated by the anticorrelation of R4570 and FWHM(H$\beta$) it is similar to EV1 found in \citet{1992boroson1, 2002boroson1} and \citet{2012xu1}.

\subsubsection{Eddington ratio and eigenvector interpretation}

In order to get more insight into the connection between the Eigenvectors and the physical properties of our sources, we calculated the Spearman rank correlation coefficients
between the Eigenvectors and the Eddington ratio ($L_{\text{bol}}$ / $L_{\text{Edd}}$). To compute the Eddington ratio we used the estimations
$L_{\text{bol}}$ = 9$\lambda$ $L_{5100}$ \citep{2000kaspi1} and $L_{\text{Edd}}$ = 1.3$\times 10^{38}$ $M_{\text{BH}}$ / $M_{\sun}$ \citep{2012xu1}.
Logarithmic mean, minimum, and maximum values for the Eddington ratio are -0.13, -0.75 and 0.48, respectively.
The correlation results are shown in Table~\ref{tab:evcorr7} and Online Figure~\ref{fig:evcorr7}.

EV2 strongly correlates with the Eddington ratio, and the correlation seems to be slightly better for the radio-loud samples. Similarly, 
\citet{2012xu1} found  a correlation between their EV1 and the Eddington ratio.
We also tried PCA with five variables - without FWHM(H$\beta$) and R4570 - and interestingly found no correlation between EV2 and Eddington ratio. I
t seems that the connection is generated by FWHM(H$\beta$) and R4570. This is supported
by the results that FWHM(H$\beta$) and R4570 correlate strongly with Eddington ratio. For FWHM(H$\beta$) -- $L_{\text{bol}}$ / $L_{\text{Edd}}$ correlation
$\rho$ = -0.71 (p=$\sim 10^{-16}$), and for R4570 -- $L_{\text{bol}}$ / $L_{\text{Edd}}$ correlation $\rho$ = 0.63 (p=$\sim 10^{-12}$).
A weak correlation between EV1 and the Eddington ratio exists only for the radio-quiet sample.

As \citet{2004grupe1} also points out, each sample has its own eigenvectors that depend on the parameters used and their range. In the earlier studies mixed samples, or samples not 
including NLS1 galaxies at all, have been used whereas we have looked at a pure NLS1 sample.
Particularly in \citet{2002boroson1} and \citet{2012xu1} the distribution of NLS1 galaxies, and also other pure samples containing just one source type, with 
respect to the two eigenvectors seems rather different from the distribution of the whole sample; distribution along EV2 is emphasized in the NLS1 group. 
This might indicate that the properties that appear in EV1 of the mixed samples, emerge in EV2 of pure samples.
This could, for example, be interpreted as a sequence of properties that change from one source type to another thus creating the EV1 in mixed samples, and which is transformed 
to a sequence of properties within a pure sample of just one source type but in a less significant role (EV2 of our NLS1 sample). EV1 of our NLS1 sample would then describe those properties 
that cause most of the differences within that source type only. We also tested our analysis by performing it with optical data only, as has been done in the earlier studies, in case 
our additional data at other wavelengths or the black hole mass might influence the result. Variables used were the SDSS $g$-band flux density, $\lambda$L5100\AA{}, F(H$\beta$), FWHM(H$\beta$), 
F([OIII]) and R4570. The values for additional optical variables were taken from \citet{2006zhou1}. This, however, did not change our results.

\begin{table}[ht]
\caption[]{Spearman rank correlations and probability values (in parentheses) for Eigenvectors 1 and 2 from PCA with seven variables, and  $\log L / L_{\text{Edd}}$ for all samples. Correlations in boldface have p$<$0.05}
\centering
\begin{tabular}{l l l}

\hline\hline
sample               & EV   &   $\log L_{\text{bol}}$ / $L_{\text{Edd}}$   \\ \hline
All                  & EV1  & -0.053 (0.605)                             \\
                     & EV2  & {\bf -0.699} (0)                           \\            
All, modified sample            & EV1  & -0.015 (0.886)                             \\            
                     & EV2  & {\bf -0.719} (0)                            \\  
Radio-loud           & EV1  & -0.184(0.247)                               \\     
                     & EV2  & {\bf -0.767} ($\sim 10^{-8}$)              \\
Radio-loud, modified sample     & EV1  & -0.136 (0.401)                             \\
                     & EV2  & {\bf -0.816} ($\sim 10^{-9}$)              \\
Radio-quiet          & EV1  & {\bf 0.301} (0.024)                         \\  
                     & EV2  & {\bf -0.598} ($\sim 10^{-6}$)              \\ \hline

\end{tabular}
\label{tab:evcorr7}
\end{table}

\onlfig{
\begin{figure*}[ht!]
\centering

  \subfloat{\includegraphics[width=0.49\textwidth]{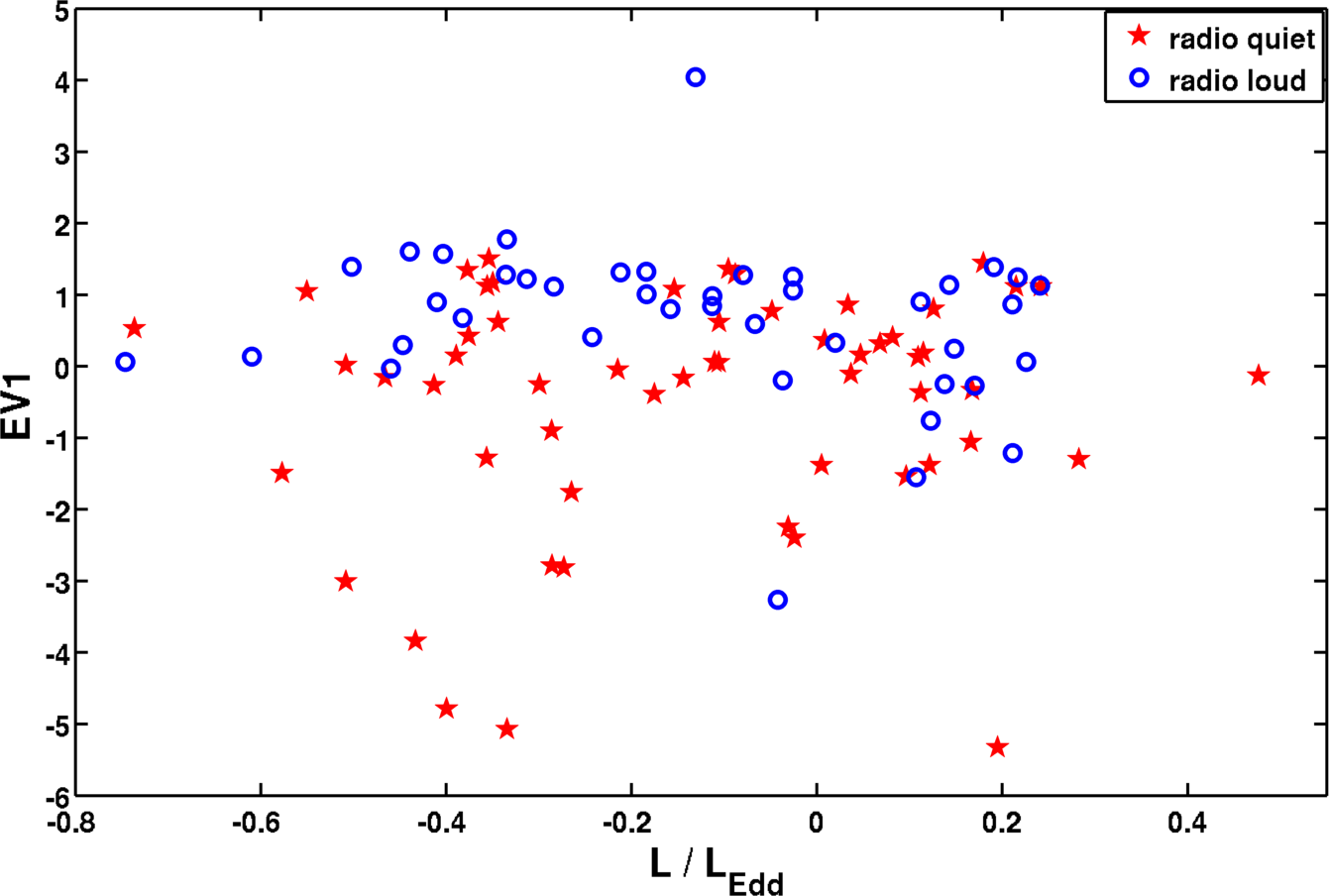}} \hspace{0.2cm}
  \subfloat{\includegraphics[width=0.49\textwidth]{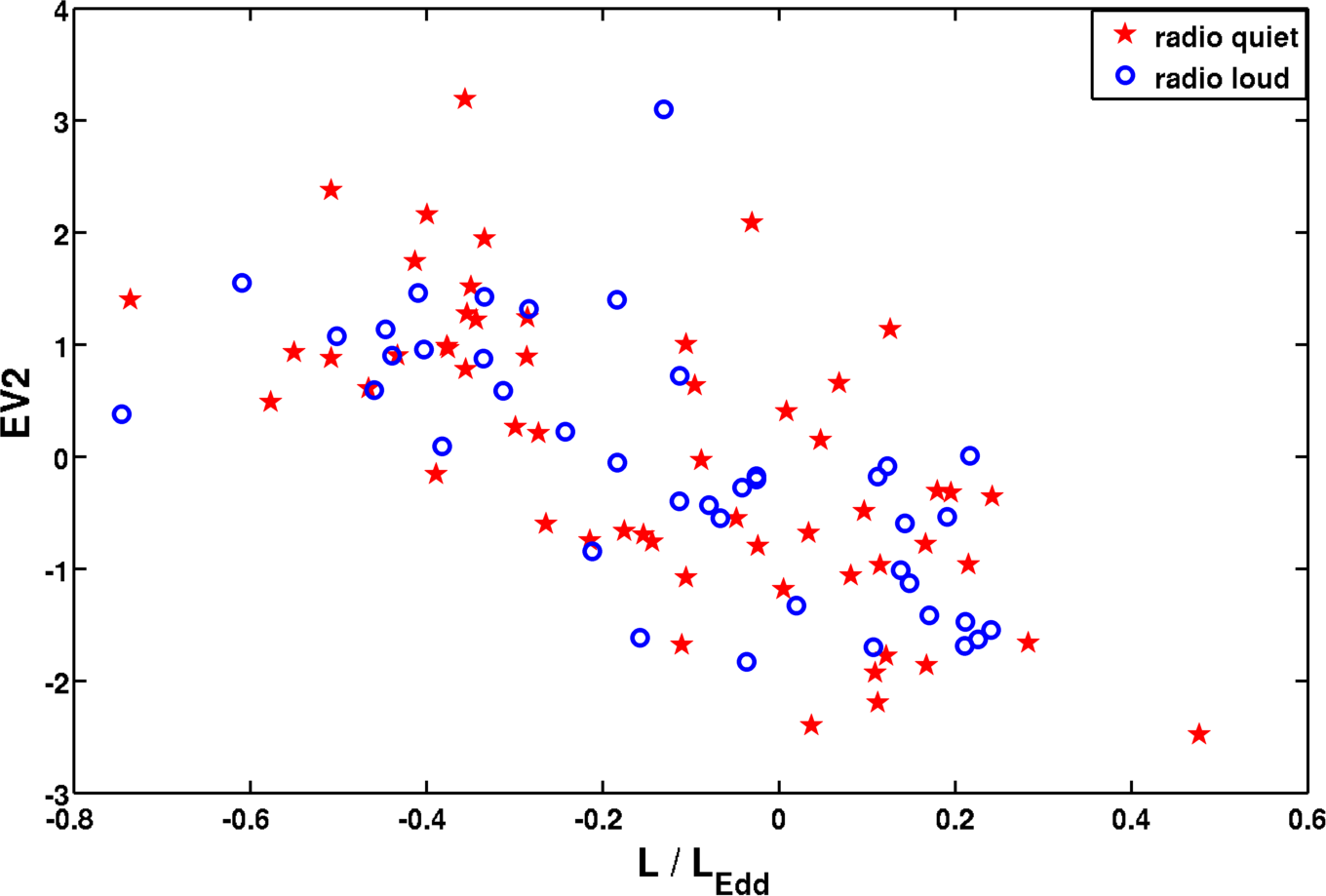}}

  \caption{Dependency between the Eddington ratio and Eigenvectors 1 and 2 from the PCA with seven variables. Radio-loud and radio-quiet subsamples are shown with different symbols and colors;
  radio-quiet: filled red stars, and radio-loud: open blue circles.}
  
\label{fig:evcorr7}
\end{figure*}
}

\subsection{$M_{\text{BH}}$ -- radio-loudness correlations}
\label{sec:mbhrlcorr}

Whether the correlation $M_{\text{BH}}$ -- $RL$ in AGN really exists has been widely studied. The results are contradictory. Some studies
suggest that the correlation exists (e.g., \citet{2000laor1, 2001lacy1, 2004mclure1, 2006metcalf1, 2011chiaberge1, 2013castignani1})
whereas others conclude that there is no correlation \citep{2002oshlack1, 2002woo2} or the correlation
at least is not clear \citep{2002ho1, 2002woo1}. In a study of 47 NLS1 galaxies \citet{2006whalen1} found an anticorrelation between $M_{\text{BH}}$ and RL.

We found that the black hole mass and the radio-loudness weakly correlate for the whole sample with Pearson's r = 0.244 (p =$\sim$10$^{-5}$)
and Spearman's $\rho$ = 0.212 (p=$\sim$10$^{-4}$). This indicates that NLS1 galaxies with more massive black holes are more likely to be able 
to lauch powerful relativistic jets and are therefore louder at radio frequencies. 

\begin{figure}[]
\centering
 \includegraphics[width=0.5\textwidth]{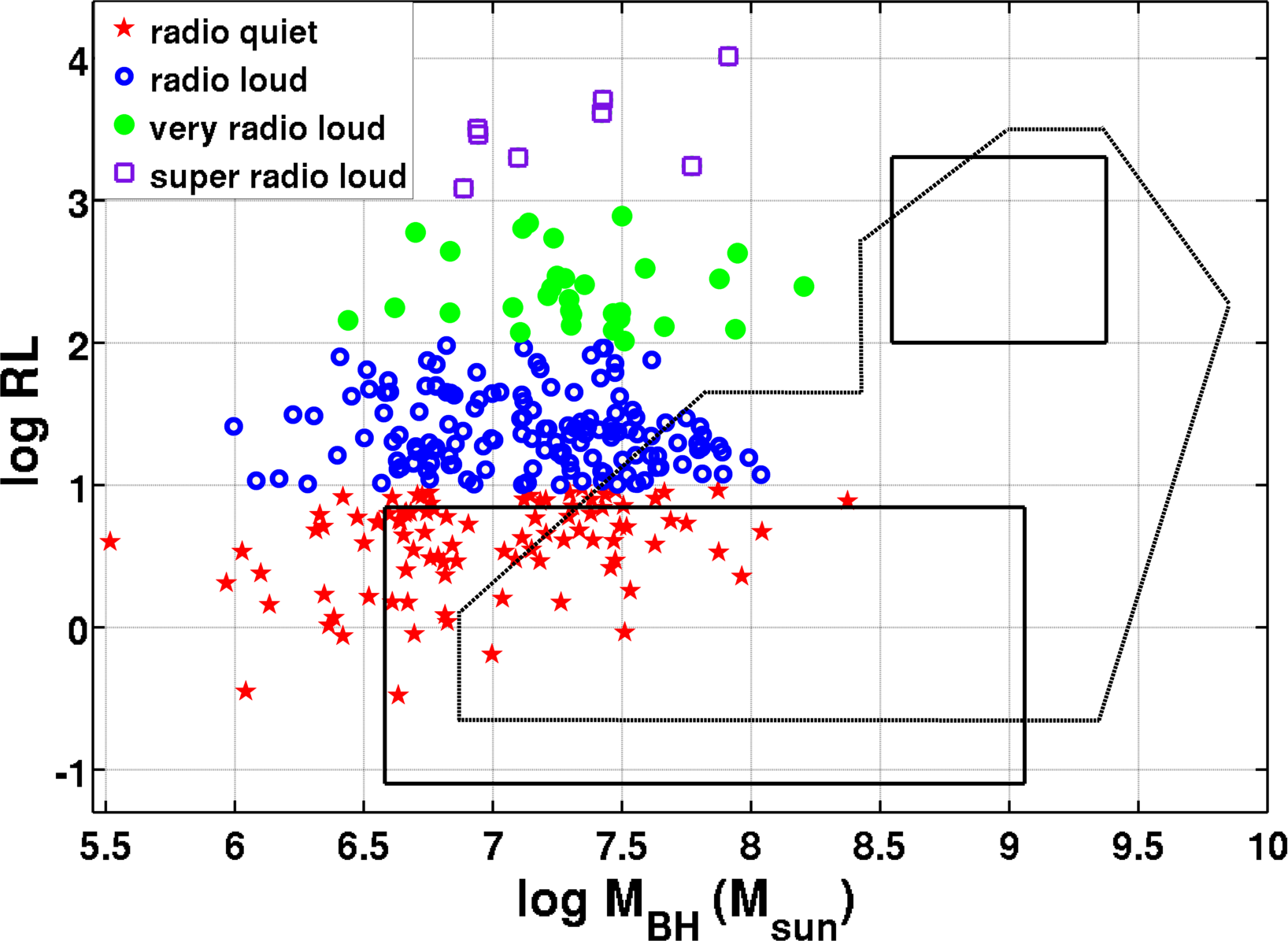}
 \caption{Dependence of radio-loudness on black hole mass. Formerly populated regions from \citet{2006komossa1} shown with solid and dashed lines. 
  Subsamples are shown with different symbols and colors; radio-quiet: filled red stars, radio-loud: open blue
  circles, very radio-loud: filled green circles, and super radio-loud: open purple squares.}
 \label{fig:laor1}
\end{figure}

\citet{2006komossa1} noticed that in the Laor diagram ($M_{\text{BH}}$ -- $RL$ \citep{2000laor1}) NLS1 galaxies are located in a region formerly
very sparsely populated with sources with rather small black hole masses (given their radio-loudness) compared to other AGN.
We constructed a Laor diagram of our whole sample (Figure~\ref{fig:laor1}). Most of our sources also lie outside the formerly 
populated regions shown in Figure 4 in \citet{2006komossa1}.

\subsection{$M_{\text{BH}}$ -- luminosity correlations}
\label{sec:mbhlcorr}

Several previous studies have not shown $M_{\text{BH}}$ -- $L_{\text{R}}$ correlation in AGN
(e.g., \citet{2002ho1, 2002woo1, 2002oshlack1, 2003snellen1, 2005woo1, 2011leontavares1, 2013park1}). \citet{2005woo1} also studied the
$M_{\text{BH}}$ -- $L_{\text{X}}$ correlation (particularly in BL Lac objects) and did not find any support for it.
In a study of radio-loud quasars \citet{2006metcalf1} found no clear $M_{\text{BH}}$ -- $L_{\text{R}}$ correlation but they did find a strong
$M_{\text{BH}}$ -- $L_{\text{O}}$ correlation.
A large number of studies find a correlation between $M_{\text{BH}}$ and $L_{\text{R}}$
(e.g., \citet{1999mclure1, 2001lacy1, 2003dunlop1, 2004mclure1, 2009bianchi1}). \citet{2009bianchi1} also found correlation between
$M_{\text{BH}}$ and $L_{\text{X}}$.

We calculated the correlations using the FIRST, SDSS $g$-band, WISE W1-band, and RASS luminosities.
The results are presented in Online Table~\ref{tab:mbh-lum} and $M_{\text{BH}}$ -- L plots in Online Figure~\ref{fig:mbhl}.
$M_{\text{BH}}$ and luminosities at all wavebands correlate well for the whole sample and all subsamples;
the correlation is strongest between $M_{\text{BH}}$ and $L_{\text{O}}$, and $M_{\text{BH}}$ and $L_{\text{IR}}$.
This suggests that the more massive the black hole the more powerful the AGN. In radio-loud NLS1 sources, where most of the emission
supposedly comes from the jet, this suggests that more massive black holes have more powerful jets. In radio-quiet sources, if radio
emission comes from supernova remnants and infrared emission from star formation, it is harder to explain why $M_{\text{BH}}$ correlates 
with $L_{\text{R}}$ and $L_{\text{IR}}$.

\onltab{
\begin{table*}[ht!]
\caption{$M_{\text{BH}}$ -- luminosity correlations for the whole sample and the subsamples. The upper value is Pearson's r
and the lower value is Spearman's $\rho$ (p-value in parentheses). Correlations in boldface have p$<$0.05}
\label{tab:mbh-lum}
\centering
\begin{tabular}{l l l l l l}

\hline\hline
\multicolumn{2}{c}{}              &  $\log M_{\text{BH}}$ --  $\log L_{\text{R}}$ &  $\log M_{\text{BH}}$ --  $\log L_{\text{O}}$ &  $\log M_{\text{BH}}$ --  $\log L_{\text{IR}}$ &  $\log M_{\text{BH}}$ -- $\log L_{\text{X}}$ \\ \hline
\multirow{2}{*}{All} & r (p)      & {\bf 0.632} ($\sim$10$^{-32}$)                      & {\bf 0.836} ($\sim$10$^{-73}$)            & {\bf 0.831} ($\sim$10$^{-69}$)             & {\bf 0.683} ($\sim$10$^{-15}$) \\      
                     & $\rho$ (p) & {\bf 0.673} ($\sim$0)                              & {\bf 0.857} ($\sim$0)                     & {\bf 0.840} ($\sim$0)                      & {\bf 0.711} ($\sim$0) \\       
\multirow{2}{*}{RQ}  & r (p)      & {\bf 0.790} ($\sim$10$^{-21}$)                      & {\bf 0.843} ($\sim$10$^{-26}$)            & {\bf 0.890} ($\sim$10$^{-30}$)             & {\bf 0.679} ($\sim$10$^{-9}$) \\      
                     & $\rho$ (p) & {\bf 0.793} ($\sim$0)                              & {\bf 0.848} ($\sim$0)                     & {\bf 0.894} ($\sim$0)                      & {\bf 0.716} ($\sim$0) \\         
\multirow{2}{*}{RL}  & r (p)      & {\bf 0.580} ($\sim$10$^{-17}$)                      & {\bf 0.823} ($\sim$10$^{-46}$)            & {\bf 0.788} ($\sim$10$^{-39}$)             & {\bf 0.669} ($\sim$10$^{-6}$) \\       
                     & $\rho$ (p) & {\bf 0.634} ($\sim$0)                              & {\bf 0.836} ($\sim$0)                     & {\bf 0.803} ($\sim$0)                      & {\bf 0.647} ($\sim$10$^{-6}$) \\        
\multirow{2}{*}{VRL} & r (p)      & {\bf 0.470} (0.003)                                & {\bf 0.748} ($\sim$10$^{-8}$)             & {\bf 0.832} ($\sim$10$^{-11}$)             & {\bf 0.704} (0.034) \\       
                     & $\rho$ (p) & {\bf 0.351} (0.029)                                & {\bf 0.671} ($\sim$10$^{-6}$)             & {\bf 0.803} ($\sim$10$^{-8}$)              & {\bf 0.700} (0.043) \\        
\multirow{2}{*}{SRL} & r (p)      & {\bf 0.768} (0.026)                                & {\bf 0.769} (0.026)                       & {\bf 0.794} (0.019)                        & 0.608 (0.584) \\          
                     & $\rho$ (p) & {\bf 0.786} (0.028)                                & {\bf 0.810} (0.022)                       & {\bf 0.833} (0.015)                        & 0.500 (1.000) \\ \hline

\end{tabular}
\end{table*}
}

\onlfig{
\begin{figure*}[ht!]
\centering

  \subfloat{\includegraphics[width=0.48\textwidth]{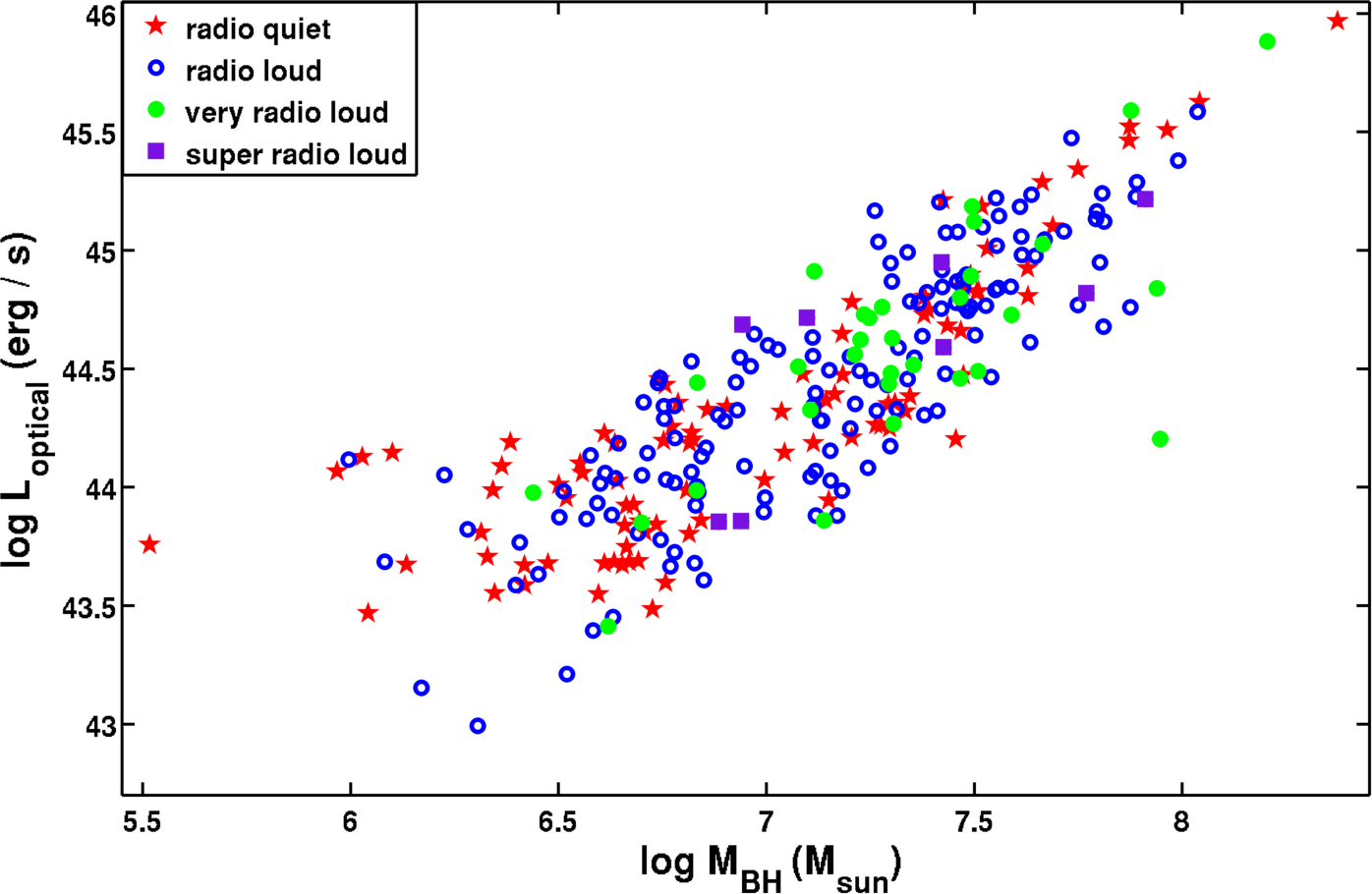}} \hspace{0.3cm}
  \subfloat{\includegraphics[width=0.49\textwidth]{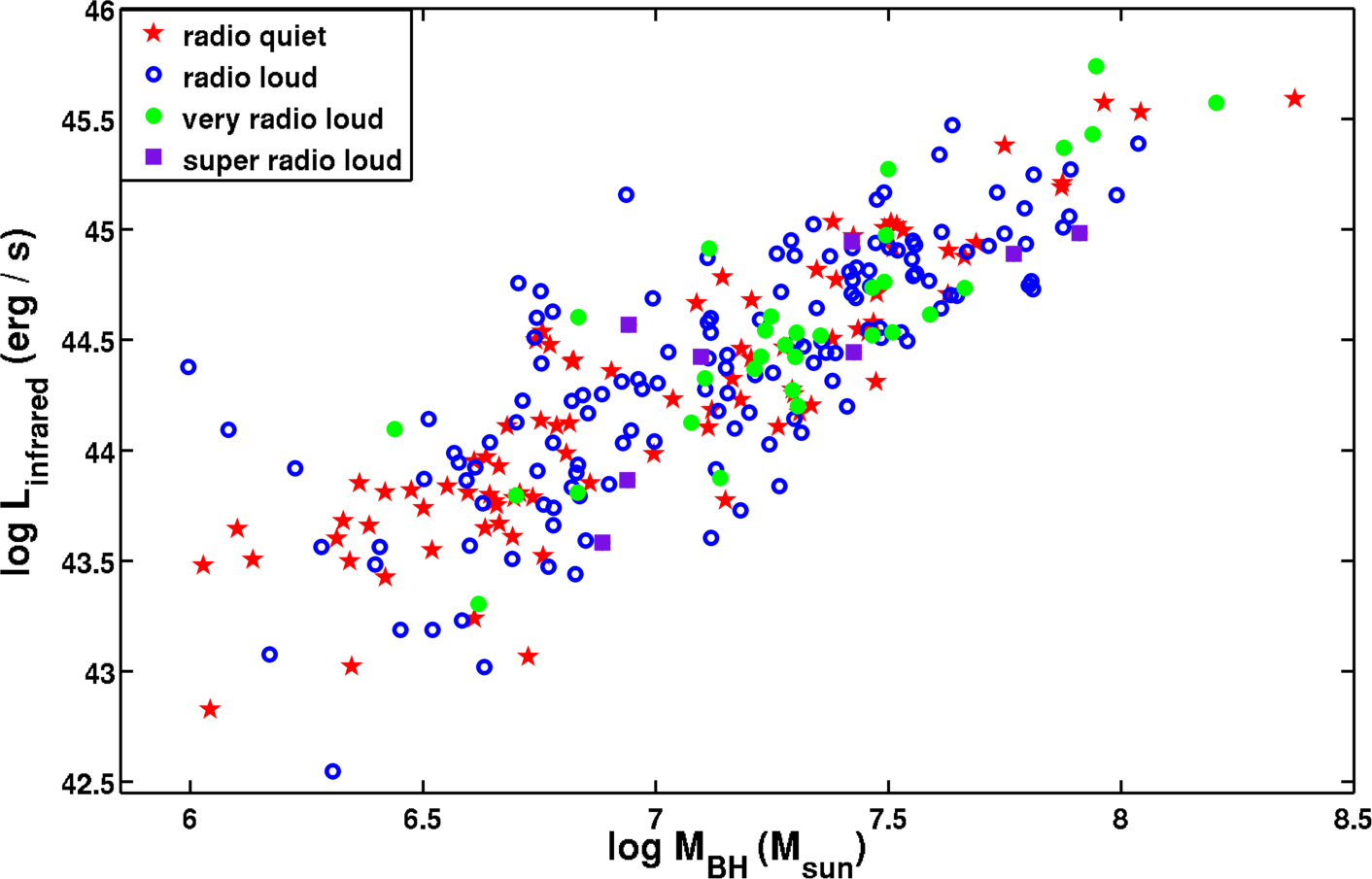}}  

  \subfloat{\includegraphics[width=0.48\textwidth]{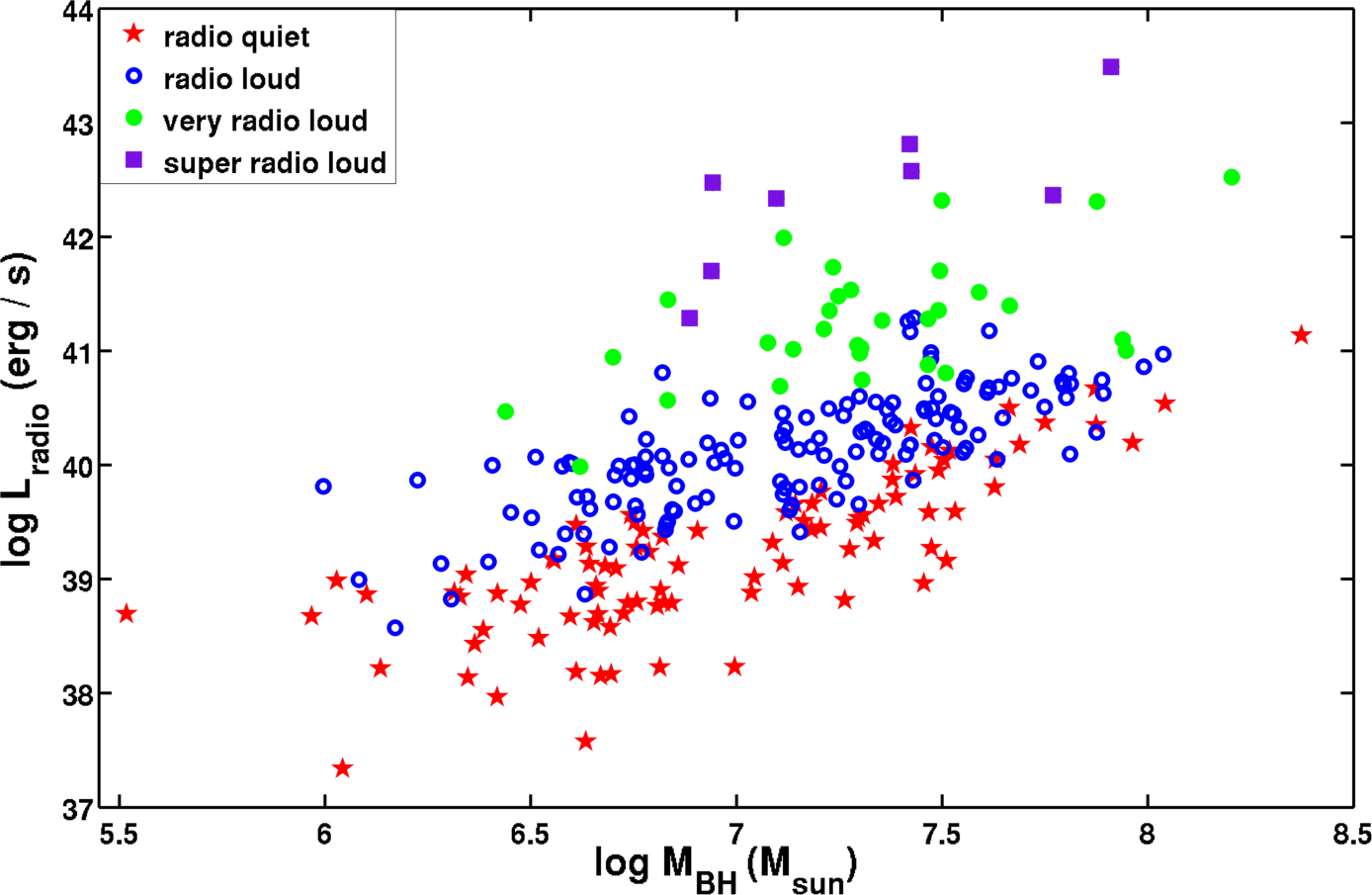}} \hspace{0.3cm}
  \subfloat{\includegraphics[width=0.49\textwidth]{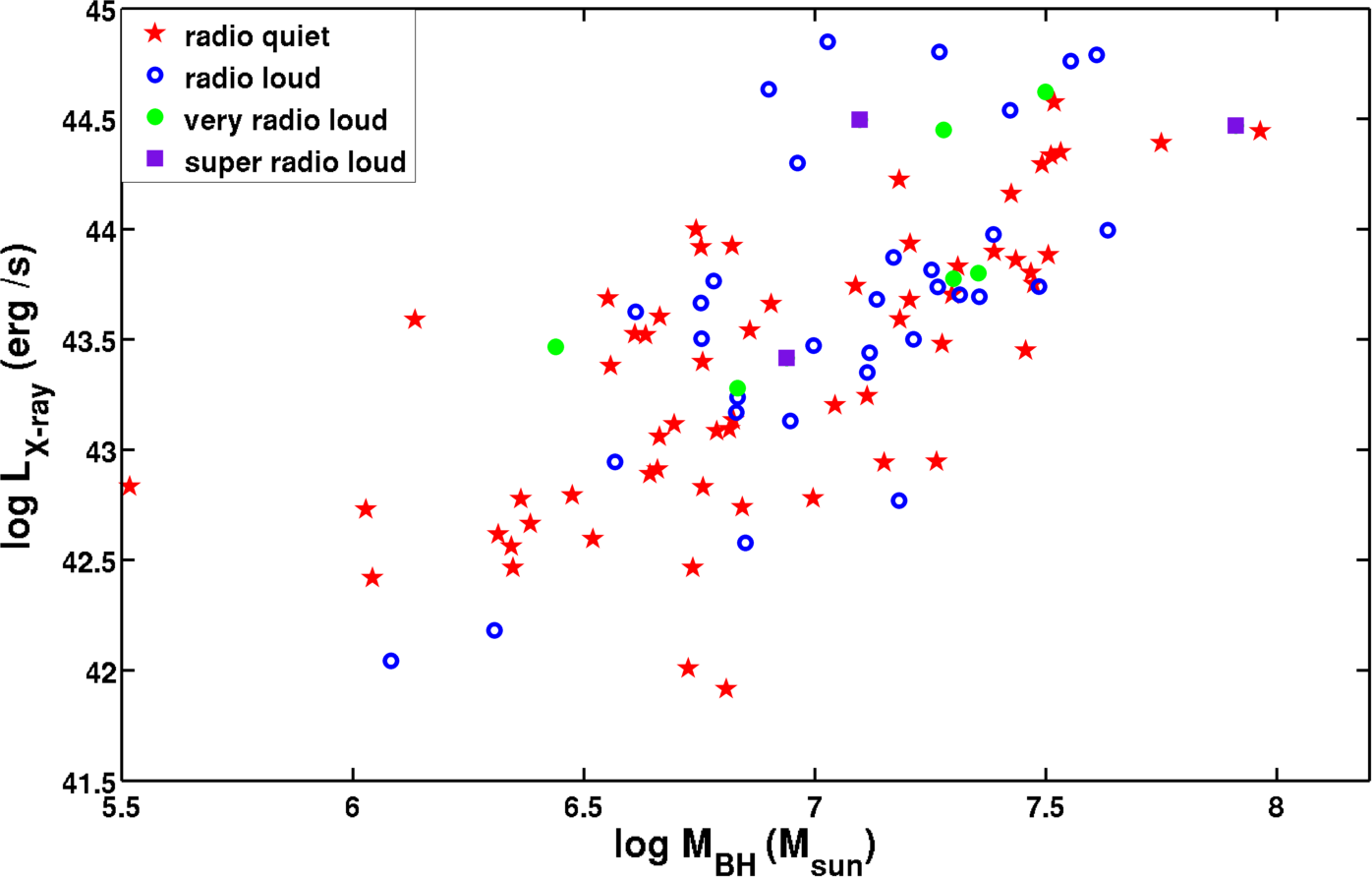}}

  \caption{Dependencies between the black hole mass and luminosities. Subsamples are shown with different symbols and colors; radio-quiet: filled red stars, radio-loud: open blue
  circles, very radio-loud: filled green circles, and super radio-loud: filled purple squares.}
 
\label{fig:mbhl}
\end{figure*}
}

\subsection{FWHM(H$\beta$) -- luminosity correlations}
\label{sec:fwhmlcorr}

The wavebands used were the same as before. The results for the 
FWHM(H$\beta$) -- $L$ correlations for the whole sample and the subsamples are presented in Online Table~\ref{tab:fwhm-lum}.
FWHM(H$\beta$) and luminosities in general correlate, although relatively weakly, for the whole sample and subsamples. Exception to this is the X-ray waveband
which does not correlate with FWHM(H$\beta$). Also there does not seem to be any correlation among the very radio-loud and super 
radio-loud subsamples. This might be due to the very limited sizes of these samples. For the whole sample the strongest correlation is 
for FWHM(H$\beta$) -- $L_{\text{O}}$. 

\onltab{
\begin{table*}[ht!]

\caption{FWHM(H$\beta$) -- luminosity correlations for the whole sample and the subsamples. The upper value is Pearson's r
and the lower value is Spearman's $\rho$ (p-value in parentheses). Correlations in boldface have p$<$0.05}

\label{tab:fwhm-lum}
\centering
\begin{tabular}{l l l l l l}
\hline\hline
\multicolumn{2}{c}{}              & FWHM --  $\log L_{\text{R}}$         & FWHM --  $\log L_{\text{O}}$   & FWHM --  $\log L_{\text{IR}}$         & FWHM --  $\log L_{\text{X}}$ \\ \hline
\multirow{2}{*}{All} & r (p)      & {\bf 0.264} ($\sim$10$^{-5}$)        & {\bf 0.305} ($\sim$10$^{-7}$)  & {\bf 0.282} ($\sim$10$^{-6}$)         & 0.024 (0.808) \\
                     & $\rho$ (p) & {\bf 0.287} ($\sim$10$^{-6}$)        & {\bf 0.302} ($\sim$10$^{-7}$)  & {\bf 0.298} ($\sim$10$^{-7}$)         & 0.045 (0.655) \\   
\multirow{2}{*}{RQ}  & r (p)      & {\bf 0.312} (0.002)                  & {\bf 0.323} (0.001)            & {\bf 0.361} ($\sim$10$^{-4}$)         & 0.059 (0.653) \\ 
                     & $\rho$ (p) & {\bf 0.303} (0.003)                  & {\bf 0.312} (0.002)            & {\bf 0.369} ($\sim$10$^{-4}$)         & 0.094 (0.476) \\   
\multirow{2}{*}{RL}  & r (p)      & {\bf 0.242} (0.001)                  & {\bf 0.281} ($\sim$10$^{-4}$)  & {\bf 0.233} (0.002)                   & -0.099 (0.536) \\  
                     & $\rho$ (p) & {\bf 0.282} ($\sim$10$^{-4}$)        & {\bf 0.290} ($\sim$10$^{-5}$)  & {\bf 0.263} ($\sim$10$^{-4}$)         & -0.083 (0.604) \\  
\multirow{2}{*}{VRL} & r (p)      & 0.111 (0.500)                        & 0.159 (0.333)                  & 0.210 (0.199)                         & 0.015 (0.970) \\ 
                     & $\rho$ (p) & 0.056 (0.735)                        & 0.162 (0.322)                  & 0.263 (0.105)                         & -0.150 (0.708) \\   
\multirow{2}{*}{SRL} & r (p)      & 0.209 (0.620)                        & 0.111 (0.793)                  & 0.182 (0.666)                         & -0.151 (0.903) \\ 
                     & $\rho$ (p) & 0.578 (0.171)                        & 0.524 (0.197)                  & 0.595 (0.132)                         & -0.500 (1.000) \\ \hline 

\end{tabular}
\end{table*}
}

\subsection{WISE blazar strip}
\label{sec:wiseblazarstrip}

\citet{2011massaro1} showed that WISE colors can be used to differentiate sources dominated by thermal or non-thermal emission on a 
[3.4]--[4.6]--[12] $\mu$m color diagram. Using a large sample of blazars they constructed the WISE $\gamma$-ray strip (WGS); 
a region populated by sources dominated by non-thermal emission. WGS makes it possible to identify new $\gamma$-ray emitters based on their WISE colors. 
We made use of this identification tool and constructed the WISE color diagram for our sources and for the whole \citet{2006zhou1} sample. 
We used the parametrizations presented in \citet{2012massaro1}. They provided parameters for two partly 
overlapping strips; one for BL Lac objects and another for flat spectrum radio quasars (FSRQs).

\begin{figure}[ht!]
  \includegraphics[width=0.5\textwidth]{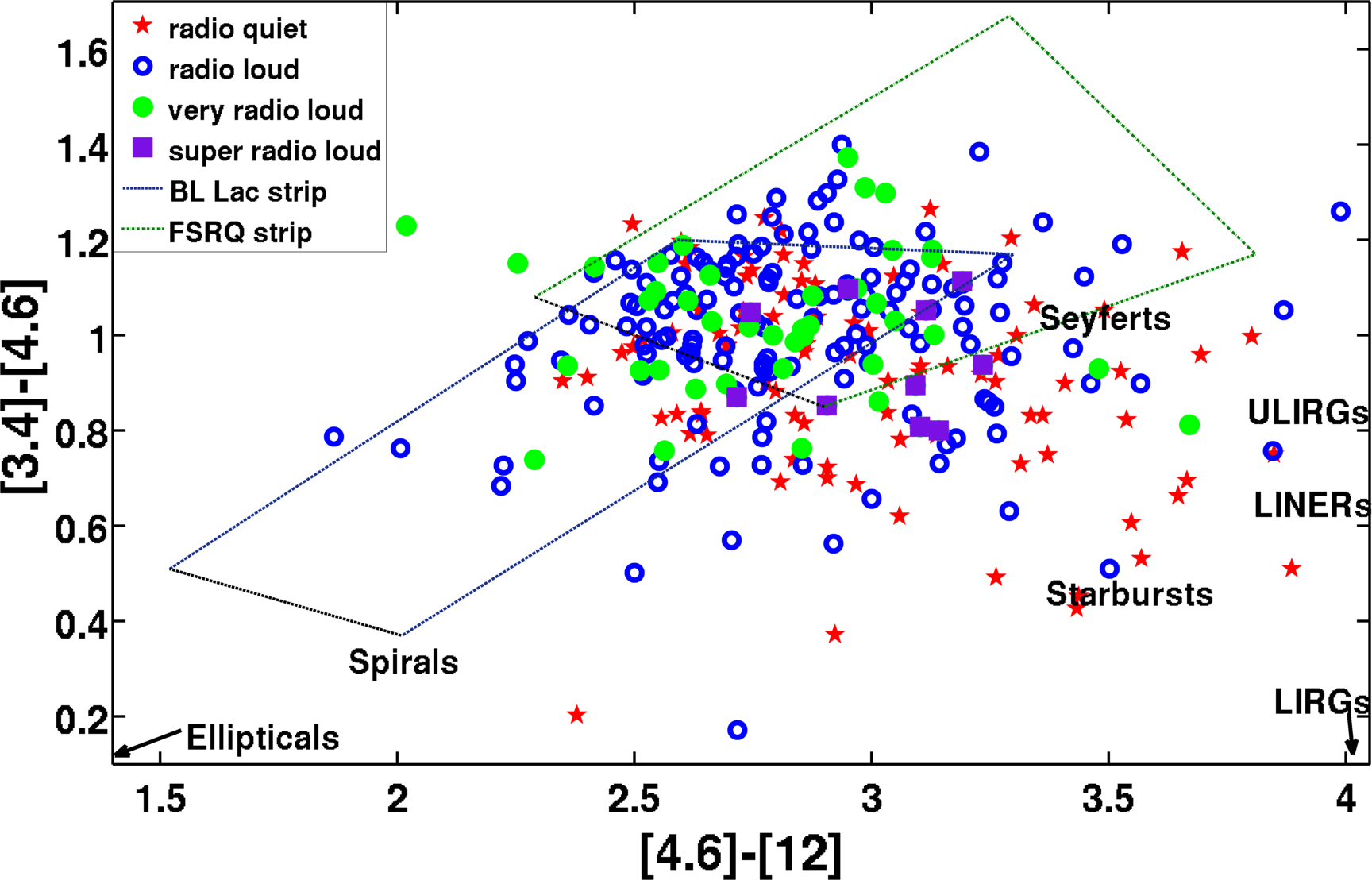}
  \caption{WISE color-color diagram of our whole sample. $\gamma$-ray strips from \citet{2012massaro1} (BL Lac strip: dark blue dashed line, and FSRQ strip: dark green dashed line)
   and approximate regions populated by other types of sources are shown. Subsamples are shown with different symbols and colors; radio-quiet: filled red stars, radio-loud: open blue
  circles, very radio-loud: filled green circles, and super radio-loud: filled purple squares. See text for details.}
  \label{fig:ourwisestrip}
\end{figure}

The WISE color diagram constructed for our sample, WGS, and approximate regions populated by other types of 
sources (ultra luminous infrared galaxies, ULIRGs; luminous infrared galaxies, LIRGs; low-ionization nuclear 
emission-line region galaxies, LINERs) are shown in Figure~\ref{fig:ourwisestrip}. 72.7$\%$ of all, 57.8$\%$ of 
radio-quiet and 79.7$\%$ of radio-loud sources are located inside the WGS. The properties (e.g., radio-loudness, 
flux densities, luminosities, and $M_{\text{BH}}$) for these sources are average compared to the whole sample. This
suggests that infrared emission is of non-thermal origin and that most of our sources should host a jet. Some of our results are 
contradictory to this (see {\bf \emph{On the origin of infrared emission}} in Section~\ref{sec:discussion}). We expected that the 
difference between radio-quiet and radio-loud sources would be bigger since radio-loud sources should host a jet whereas
radio-quiet should not. The reason for so many radio-quiet sources lying inside the WGS is unclear, but there 
are a few possible explanations. It might be that for some, yet unclear reason WGS does not apply for NLS1 galaxies. This might be due to the 
enhanced star formation in the circumnuclear regions of NLS1 galaxies because the star formation process is 'boosting' the infrared colors. 
Another possibility is that these sources really host a jet, but it is not powerful enough to dominate the whole electromagnetic spectrum.

1348 out of 1943 WISE detected sources from \citet{2006zhou1} lie inside the WGS. This is nearly the same percentage (69.4\%) as in our sample which indicates
that our sample represents the whole sample well.

\section{Discussion}
\label{sec:discussion}

{\bf \emph{On the origin of infrared emission.}} Linear slope fitting suggests that infrared emission from 
radio-quiet and radio-loud sources has a different origin, supposedly thermal in radio-quiet and non-thermal in 
radio-loud sources. Multiwavelength correlations indicate that infrared emission in radio-quiet sources might be thermal whereas WGS 
indicates that it is non-thermal. The WGS also suggests that the origin in radio-loud sources is non-thermal. On the other hand 
the multiwavelength correlation results for radio-loud sources suggest that infrared emission is of thermal origin. Principal component analysis
suggests that the origin of the infrared emission in both radio-loud and radio-quiet sources is of thermal origin. Thus the source, or sources, of the 
infrared emission in NLS1 galaxies remains unclear. It is possible that non-thermal, reradiated (from torus), and star forming 
processes all contribute to the infrared emission. More infrared observations of large NLS1 samples are needed in order to 
study the origins of the infrared emission in more detail.

{\bf \emph{Parent population.}} The results for the K-S test for luminosities, redshift, and black hole masses indicate that the parent population for 
the radio-loud and radio-quiet NLS1 sources is different. Still they have many similar properties, for example, the optical spectrum, which suggests that 
they are from the same or from very similar parent populations. To get an insight into the differences between the radio-loud and radio-quiet populations, 
a proper characterization of the host galaxies of radio-loud and radio-quiet sources is needed to see whether the differences might be related to the host 
galaxy type. Differences in black hole masses between the radio-quiet and radio-loud sources might also suggest an evolutionary line.

In our sample, radio-quiet sources tend to lie closer than radio-loud sources. This may also be related to AGN evolution;
AGN were more numerous and luminous in the past. AGNs with more massive black holes -- and thus higher luminosities -- evolved 
first and were highly active in the past. They evolved to be less luminous or entirely quiescent sources, e.g., normal
non-active galaxies. Low-luminosity sources with lower mass black holes emerged later and are much more numerous in the present
Universe \citep{2012beckmann1}. Also the significant $z$ --  $RL$ correlation
(Pearson's r = 0.474, p=$\sim10^{-18}$ and Spearman's $\rho$ = 0.497, p=$\sim10^{-19}$) supports these results; radio-quiet
sources tend to lie closer than radio-loud.

In the current unification models differences between Type 1 and 2 Seyfert galaxies are explained with orientation and obscuration effects. 
In these models, Seyfert 1 galaxies are viewed pole-on, unobstructed, while Seyfert 2 galaxies 
are viewed edge-on through obscuring matter in the accretion disk (e.g. \citet{1983miller1} and \citet{1993antonucci1}). 
However, evidence against the simple unification model of Seyfert 1 and 2 galaxies has lately accumulated. Several studies
have shown that approximately 50$\%$ of Seyfert 2 galaxies do not show a hidden broad-line region (HBLR) in their polarized 
optical spectra. This suggests that not all Seyfert 2 galaxies harbor a nucleus similar to Seyfert 1 galaxies. It seems that 
non-HBLR Seyfert 2 galaxies accommodate weaker nuclei that do not exhibit typical BLRs.
HBLR and non-HBLR Seyfert 2 galaxies have also other differences such as luminosity and accretion rate 
(e.g., \citet{2001tran1, 2003tran1, 2011wu1, 2012marinucci1, 2013yu1}). 
\citet{2006zhang1} studied NLS1 and non-HBLR Seyfert 2 galaxies, and suggested that they can be unified based on orientation; 
non-HBLR Seyfert 2 galaxies are NLS1 galaxies viewed edge-on. This result is, however, debatable since the sources have many 
unexplained differences, for example, non-HBLR Seyfert 2 galaxies do not show Fe II emission lines \citep{2013yu1}.
Furthermore, \citet{2008decarli1} studied the properties of NLS1 and BLS1 galaxies. They showed that when assuming a disk-like BLR and pole-on
orientation of NLS1 galaxies, some of the observed differences between NLS1 and BLS1 galaxies can be explained.

AGN unification models are currently under constant revision; the question how NLS1 galaxies fit in remains open. 

\section{Summary}
\label{conclusion}

In this study the aim was to further our understanding about Narrow-Line Seyfert 1 galaxies, which are a new
class of $\gamma$-ray emitting AGN. We addressed this issue by studying emission processes and properties of NLS1 galaxies,
and also studied how these properties are connected with other AGN properties. To this end, we have compiled the largest multiwavelength database 
of NLS1 galaxies so far. This data set should be useful in the future as it permits easy identification and comparison of properties of 
new $\gamma$-ray NLS1 sources to be detected by Fermi or other facilities.

The main results of this study are: 

\begin{enumerate}
      \item The distributions of radio-quiet and radio-loud NLS1 galaxies in redshift, luminosity, and black hole mass are different. Radio-quiet sources also tend to lie closer than radio-loud sources.
      \item NLS1 sources with more massive black holes are more likely to be able to lauch a powerful relativistic jet.
      \item Multiwavelength correlations suggest that radio-loud sources host a jet which is the predominant sources of the radio, optical, 
            and at least partially also X-ray emission. The origin of the infrared emission remains unclear.
      \item Radio-quiet sources do not host a jet or the jet is very weak. Radio and infrared emission are more likely
            to originate from star formation processes, and optical and X-ray emission from the inner parts of the AGN.
\end{enumerate}

While the results of this study mainly confirm what is already known of NLS1 galaxies, they also serve as a comprehensive starting point for further studies, 
for example, at high radio and infrared frequencies where the information so far is scarce.
Open questions in need of more investigation include, for example, the origin of the infrared emission in NLS1 galaxies and the differences between 
radio-quiet and radio-loud sources (i.e. the question of the parent population). Simultaneous multiwavelength observations of especially radio-loud NLS1 
sources should be very useful in modeling and understanding their emission properties. Studies of NLS1 galaxies may also provide us with a further look 
at AGN evolution and activity in general, and the evolution of relativistic jets.

\begin{acknowledgements}

We are grateful to K. I. I. Koljonen for his help with PCA. This research has made use of the NASA/IPAC Extragalactic Database (NED) which is operated by the Jet Propulsion Laboratory, 
California Institute of Technology, under contract with the National Aeronautics and Space Administration. 

The National Radio Astronomy Observatory is a facility of the National Science Foundation operated under cooperative agreement by Associated Universities, Inc.

This publication makes use of data products from the Wide-field Infrared Survey Explorer, which is a joint project of the University of California,
Los Angeles, and the Jet Propulsion Laboratory/California Institute of Technology, funded by the National Aeronautics and Space Administration.

Funding for the Sloan Digital Sky Survey (SDSS) has been provided by the Alfred P. Sloan Foundation, the Participating Institutions, 
the National Aeronautics and Space Administration, the National Science Foundation, the U.S. Department of Energy, the Japanese Monbukagakusho, 
and the Max Planck Society. The SDSS Web site is http://www.sdss.org/.

The SDSS is managed by the Astrophysical Research Consortium (ARC) for the Participating Institutions. The Participating Institutions are 
The University of Chicago, Fermilab, the Institute for Advanced Study, the Japan Participation Group, The Johns Hopkins University, 
the Korean Scientist Group, Los Alamos National Laboratory, the Max-Planck-Institute for Astronomy (MPIA), the Max-Planck-Institute for Astrophysics (MPA), 
New Mexico State University, University of Pittsburgh, University of Portsmouth, Princeton University, 
the United States Naval Observatory, and the University of Washington.

This research has made use of the ROSAT All-Sky Survey data which have been processed at MPE.

\end{acknowledgements}

\bibliographystyle{aa}
\bibliography{artikkeli.bib}

\end{document}